\documentclass[aps,prl,twocolumn,amsmath,amssymb,amsfonts,nofootinbib,long,floatfix]{revtex4}
\usepackage{graphicx,epsfig,latexsym,bm}

\newcommand\eV{\mbox{eV}}

\newcommand\GeV{\mbox{GeV}}
\newcommand\pc{\mbox{pc}}
\newcommand\kpc{\mbox{kpc}}
\newcommand\Mpc{\mbox{Mpc}}
\newcommand\G{\mbox{G}}
\newcommand\A{\mathbf{A}}
\newcommand\B{\mathbf{B}}
\newcommand\F{\mathbf{F}}
\newcommand\x{\mathbf{x}}
\newcommand\y{\mathbf{y}}
\newcommand\kk{\mathbf{k}}
\newcommand\ee{{\boldsymbol \varepsilon}}
\newcommand\mPl{m_{\rm Pl}}

\newcommand\E{\mathbf{E}}
\newcommand\PP{\mathbf{P}}
\newcommand\X{\mathbf{X}}
\newcommand\D{\mathbf{D}}
\newcommand\HH{\mathbf{H}}
\newcommand\M{\mathbf{M}}

\begin{document}

\title{Lorentz-violating inflationary magnetogenesis}

\author{Leonardo Campanelli$^{1}$}
\email{leonardo.campanelli@ba.infn.it}
\affiliation{$^1$Dipartimento di Fisica, Universit\`{a} di Bari, I-70126 Bari, Italy}

\date{\today}


\begin{abstract}
A non-conformally invariant coupling between the inflaton and the photon in the minimal
Lorentz-violating standard model extension is analyzed. For specific forms of the
Lorentz-violating background tensor, the strong-coupling and backreaction problems of
magnetogenesis in de Sitter inflation with scale $\sim 10^{16} \GeV$ are evaded,
the electromagnetic-induced primordial spectra of (Gaussian and non-Gaussian) scalar and tensor
curvature perturbations are compatible with cosmic microwave background observations, and the
inflation-produced magnetic field directly accounts for cosmic magnetic fields.
\end{abstract}




\maketitle


\section{I. Introduction}

Coherent magnetic fields as strong as $B \sim 10^{-6} \G$ have
been detected in any type of galaxies and in galaxy clusters
(for reviews on cosmic magnetic fields,
see~\cite{Review1,Review2,Review3,Review4,Review5,Review6,Review7,Review8,Review9}).
Their origin is still an open issue and is puzzling to the point that
``cosmic magnetism'' should be considered one of the biggest
mysteries in cosmology.

Nowadays, what it is clear enough is that seed magnetic fields
present prior to galaxy formation can be
amplified by protogalaxy collapse and magnetohydrodynamic turbulence effects and then,
at least in principle, they can reproduce the
properties of presently-observed galactic fields.

Indeed, it has been recently pointed out~\cite{Pakmor} (see also references therein) that
a small-scale dynamo could exponentially amplify small-scale seed magnetic fields during the process of
galactic disk formation. Successively, differential rotation
of the newly formed galactic disk would order the chaotic
field resulting from the small-scale dynamo in such a way to reproduce the main features
of the observed galactic magnetic fields.
This mechanism would explain galactic magnetism if a sufficiently strong seed field is present prior to galaxy formation
but leave substantially unanswered the question of the presence of
strong magnetic fields in clusters of galaxies.

A plethora of mechanism acting in the early Universe have been proposed to produce seed
fields since Fermi's proposal of the existence of cosmic magnetic fields back in 1949~\cite{Fermi}.

Promising candidates are those mechanism operating during inflation since
inflation-generated fields can be correlated on super-horizon scales, and
then their comoving correlation length can be as large as the
galactic one.
If magnetic fields are created after inflation, instead, their
correlation length cannot exceed the dimension of the horizon at the time of generation, so
that they are correlated on scales generally much smaller than the
characteristic scale of the observed cosmic fields.

Since standard Maxwell electromagnetism in a Friedmann-Robertson-Walker universe
is invariant under conformal transformations, magnetic fields cannot be generated during inflation,
as a consequence of the well-known ``Parker theorem''~\cite{Birrell-Davies,Parker-Toms}.
For this reason, all inflationary models proposed in the
literature repose on the breaking of conformal invariance of (standard) electrodynamics.

Turner and Widrow~\cite{Turner-Widrow} analyzed the consequences
of adding, to the Maxwell Lagrangian, nonstandard conformal-breaking gravitational
couplings of the photon.

Ratra~\cite{Ratra}, instead, introduced a nonstandard conformal-breaking coupling between the scalar
field $\phi$ responsible for inflation (the inflaton) and the electromagnetic field.

After these two seminal papers on the generation of large-scale magnetic fields
at inflation, many other mechanisms have been proposed,
most of which introduces nonstandard photon couplings to break conformal invariance
(see, e.g.,~\cite{G2,G3,G4,G5,G6,G7,G8,G9,G10,G11,G12,G13,G14,G15,G16,G17,G18,G19,
G21,G22,G23,G24,G25,G26,G27,G29,G30,G31,G32,G34,Campanelli2,G34bis,G35,G35bis,G36,G37}).

There are, however, three mechanisms proposed in the literature that
work without resorting to nonstandard physics.

Dolgov~\cite{Dolgov} argued that the well-know conformal anomaly in quantum field
theory in curved spacetime induces a breaking of conformal invariance of standard
electrodynamics, which in turn stimulates the generation of strong, large-scale
magnetic field at inflation.

Barrow and Tsagas~\cite{Barrow}  (see, also,~\cite{Barrow1} and ~\cite{Barrow2})
showed that, within the framework of conventional electromagnetism, astrophysically
interesting magnetic fields can be generated if one assumes, contrarily to what
previously assumed in the literature of cosmic magnetic fields, that the spatial curvature
of the Universe is nonzero and compatible with astrophysical observations.

The author pointed out in~\cite{Campanelli} (see, also,~\cite{Campanelli5})
that the process of renormalization of inflationary quantum magnetic fluctuations
naturally breaks conformal invariance giving, as a result, a strong, scale-independent
today magnetic field.

Recently enough, however, a potential problem for inflationary mechanisms of magnetogenesis
has been pinpointed by Demozzi, Mukhanov, and Rubinstein~\cite{Demozzi}, and it is now known
as the ``strong-coupling problem''. The problem
consists in the fact that the full electrodynamics theory, including both the nonstandard couplings of the photon with other fields
(as the inflaton) and the standard one with conserved external currents, must always be in a weak-coupling
regime, in order to have reliable results.
This problem, when combined with the so-called ``backreaction problem'', which appears in the theory when the
inflation-produced electromagnetic field appreciably back-reacts on the inflationary dynamics,
excludes all the models of inflationary magnetogenesis based on nonstandard physics.

After this work~\cite{Demozzi}, only three scenarios for inflationary magnetogenesis
have been suggested in which both the strong-coupling and the backreaction problems are avoided.
(Such problems are successfully evaded also in a magnetogenesis model proposed by Membiela~\cite{Membiela}.
In such a model, however, the background cosmology is given by a nonstandard bouncing cosmological model
instead of standard inflation.)

Ferreira, Jain, and Sloth~\cite{Ferreira} considered a magnetogenesis scenario {\it \`{a} la} Ratra where
the inflaton is kinetically coupled to the photon and where a low scale inflation is followed by a prolonged
reheating phase dominated by a stiff fluid.

Caprini and Sorbo~\cite{Caprini} proposed a generalization of the Ratra-like model,
where both a kinetic and an axion-like coupling are present.

Very interesting is the scenario recently proposed by Tasinato~\cite{Tasinato}.
A (nonstandard) derivative interaction between fermion fields (which give rise to the external currents)
and a scalar field (which is kinetically coupled to the photon and amplifies electromagnetic vacuum fluctuations)
``renormalizes'' the electric charge during inflation in such a way that the theory is always in the
weak-coupling regime. (For possible problems that could arise in this scenario, see~\cite{Ferreira2}).

Beside the strong coupling and the backreaction problems,
there is another possible problem, hereinafter referred to as the ``curvature perturbation problem'',
first pointed out by Barnaby, Namba, and Peloso~\cite{Barnaby}.
It consists in the fact that inflationary electromagnetic fields
generate both scalar (Gaussian and non-Gaussian) and tensor curvature perturbations that
could be in conflict with recent observations of cosmic microwave background (CMB)
anisotropies.

In this paper, we discuss a generalization of the Ratra model where the inflaton $\phi$ is kinetically
coupled to the photon through a Lorentz-violating coupling of the form
\begin{equation}
\label{intro}
f(\phi) (\mathcal{L}_M + \mathcal{L}_{LV}),
\end{equation}
where $f$ is a generic positive-defined function, $\mathcal{L}_M$ is the standard Maxwell Lagrangian,
while $\mathcal{L}_{LV}$ contains all Lorentz-violating terms that involve the photon field
and that are implemented by external background tensors.

The motivation behind the investigation of possible effects of Lorentz-violation
in inflationary magnetogenesis is that in some theories of quantum gravity,
such as loop quantum gravity~\cite{Gambini} and string theory~\cite{Kostelecky3},
the breakdown of Lorentz symmetry is expected to take place
around the Planck scale, and so before the beginning of inflation.

Working in the weak-coupling regime, we will show that strong, scaling-invariant, magnetic fields
can be created without back-reacting on the inflationary dynamics and without generating
curvature perturbations in conflict with CMB results.
This is possible if the external tensors, which represent new degrees of freedom with respect to the Ratra model,
assume specific (fine-tuned) forms that assure that the electric part of the electromagnetic energy-momentum tensor
(which would give rise to the the three aforementioned problems) is vanishing during inflation.

The paper is organized as follows.
In section II, we discuss the characteristics (intensity and correlation length), that
a comoving cosmic magnetic field must have in order to explain the magnetic fields
detected in galaxies and clusters of galaxies.
In section III, we briefly review, in the context of the Ratra-like model,
the strong-coupling and backreaction problems in inflationary magnetogenesis.
In sections IV and V, we introduce and quantize our model of magnetogenesis based on Lorentz-violating
couplings between the inflaton and the photon.
In section VI, we derive the conditions under which the inflation-produced electromagnetic field
does not appreciably back-react on the inflationary dynamics.
In section VII, we evolve the produced magnetic field from the end of inflation until today.
In section VIII, we discuss additional constraints that could be eventually imposed on the
inflation-produced electromagnetic energy-momentum tensor.
In section IX, we calculate the spectrum, bispectrum, and trispectrum of the scalar curvature
perturbations and estimate the spectrum of the tensor modes generated by the electromagnetic field,
and compare them to the current bounds derived by the Planck mission.
In section X, we discuss our results.
Finally, in section XI, we draw our conclusions.

\section{II. Seed magnetic fields}

Magnetic fields have been detected in all types of galaxies with
intensities of order $\mu\G$. Galaxies at high redshift (still in the process of being
formed) and irregular galaxies do not possess structured magnetic fields, while magnetic
fields in fully formed galaxies, such as spiral or barred galaxies, typically trace the
large-scale structure of galaxies~\cite{Review1,Review5}.

These observations could be explained if a sufficiently intense large-scale magnetic field
were present prior to galaxy formation. In this case, and due to the high conductivity of
the protogalactic plasma, the magnetic field would remain frozen in the plasma and its final
spatial configuration would reflect that of the galaxy. This rearrangement of the structure
(not amplification) of the magnetic field could easily be realized by a galactic dynamo action,
whose efficiency in reorganizing the primordial field is subjected to only this condition: that
the comoving magnetic correlation length be greater than about $100 \pc$~\cite{Review1}.
Moreover, due to the Alfv\'{e}n theorem (see, e.g.,~\cite{Review3}), a frozen-in magnetic field is
amplified by a factor of $[\rho_{\rm gal}/\rho_m(t)]^{2/3}$ and its correlation decreased by
$[\rho_{\rm gal}/\rho_m(t)]^{1/3}$ during protogalactic collapse~\cite{Turner-Widrow}. Here,
$\rho_{\rm gal}$ and $\rho_m(z) \propto (1+z)^3$ are, respectively, the galactic and cosmic
matter densities, and for typical galaxies $\rho_{\rm gal}/\rho_m(t_0) \sim 10^6$ at the
present cosmic time $t_0 \simeq 4 \times 10^{17} s$.
Therefore, a comoving seed field $B_0 \sim 10^{-10} [(1+z_{\rm gal})/(1+z_{\rm ta})]^2 \G$,
correlated on a comoving scale greater than
$\lambda_B \sim 10 [(1+z_{\rm ta})/(1+z_{\rm gal})] \, \kpc$, could explain the galactic
magnetism. The redshift-dependent factors come from the fact that between the turn-around
redshift  $z_{\rm ta}$ (when the protogalactic collapse begins) and the galaxy redshift
$z_{\rm gal}$, 
a frozen-in primordial magnetic field is decoupled from the Hubble flow and does not evolve
adiabatically. Typically, $z_{\rm ta} \sim few \: $tens, while $z_{\rm gal}$ ranges from 0
to $few$~\cite{Review2}. 

The observation of galaxy clusters reveals the presence of intracluster
large-scale-correlated $\mu \G$ magnetic fields. The intensity of such fields rise to tens
of $\mu \G$ in the cluster cores, but this can be probably ascribed to fast-acting dynamo
mechanisms due to cluster cooling flows~\cite{Review1}.

Numerical simulations~\cite{Dolag et al} of cluster formation starting at redshift
$z_{\rm ta} = 15$ have shown that a $few \times 10^{-10} \G$ seed field is processed by
magnetohydrodynamic effects in such a way to reproduce the observed magnetic Faraday rotation
maps of clusters at low redshifts ($z_{\rm cl} \simeq 0$). It has also be found that the
initial magnetic field correlation properties are inessential to the final result, although
the scale of the initial magnetic field fluctuations was limited by the resolution length of
order $100 \kpc$~\cite{Dolag}.
The overall amplification of a factor $few \times 10^3$ is explained as a Alfv\'{e}n
frozen-flux effect of $[\rho_{\rm cl}/\rho_m(t_0)]^{2/3} \sim 10^2$ during cluster
collapse [since, typically, $\rho_{\rm cl}/\rho_m(t_0) \sim 10^3$], plus an
amplification of a factor of $few \;$tens, probably due to a Kelvin-Helmholtz instability
of the intracluster plasma flows~\cite{Dolag et al}. Therefore, a comoving seed field
$B_0 \sim 10^{-10} (1+z_{\rm ta})^{-2} \G$, correlated on a comoving scale greater than
$\lambda_B \sim 100 \, (1+z_{\rm ta}) \, \kpc$, could explain cluster magnetic fields.

Roughly speaking, then, in order to explain both galactic magnetism and galaxy cluster
magnetic fields, it suffices to have a comoving seed magnetic field such that~\cite{Campanelli}
\begin{eqnarray}
\label{Bseed}  && 10^{-13} \G \lesssim B_0 \lesssim few \times 10^{-12} \G, \\
\label{xiseed} && \lambda_B \gtrsim few \times \Mpc.
\end{eqnarray}
{\it Limits on primordial magnetic fields.} -- If cosmic magnetic fields
are relics from inflation, they could modify the standard evolution of the
universe in radiation and matter eras. However, this is not the case, since the present
limits on primordial magnetic fields do not exclude the existence of large-scale magnetic fields
as strong as those in Eq.~(\ref{Bseed}). Indeed, the most
significant limits on large-scale cosmic magnetic fields
come from big bang nucleosynthesis analyses,
$B_0 \lesssim 1 \times 10^{-6} \G$~\cite{Cheng et al,BBN,Kahniashvili2},
data on large scale structures,
$B_0 \lesssim few \times 10^{-9} \G$~\cite{Kahniashvili,Kahniashvili2},
studies of CMB radiation,
$B_0 \lesssim few \times 10^{-9} \G$~\cite{Giovannini 2,Planck3,ellipsoidal,Kahniashvili2,Kahniashvili3},
studies of the ionization history of our Universe,
$B_0 \lesssim 10^{-9} \G$~\cite{ionization},
Faraday rotation maps of distant quasars,
$B_0 \lesssim 10^{-11} \G$~\cite{Kronberg et al, Kronberg 2},
and blazar observations,
$B_0 \gtrsim 7 \times 10^{-14} \G$~\cite{Neronov-Vovk,Tavecchio et al 1,Tavecchio et al 2},
where the last lower limit refers to the less conservative bound from the blazar
1ES 0229+200~\cite{Tavecchio et al 2}.
It is interesting to observe that the upper limit from Faraday rotation maps and the
lower limit from blazar observations are a just few times outside the interval of
$B_0$ in Eq.~(\ref{Bseed}). Narrowing the above limits
could then eventually reveal the primordial nature of cosmic magnetic fields.

\section{III. Strong coupling and backreaction in inflationary magnetogenesis}

Let us now discuss, in some detail, two requirements that have to be imposed
on any magnetogenesis mechanism operating during inflation. We focus our attention
to the ``standard'' kinetically coupled scenario for magnetogenesis,
where the inflaton field $\phi$ is coupled to the standard kinetic Maxwell term
via a generic coupling $f(\phi)$. This represents an extended version of the
model proposed by Ratra~\cite{Ratra},
where $f(\phi) \propto e^{\alpha \phi}$, with $\alpha$ being a constant.

The first requirement, that the full theory,
namely when including conserved external currents, must be in a weak-coupling regime, has been
discussed only recently in~\cite{Demozzi}.

The second requirement, namely that the inflation-produced
electromagnetic field must not appreciably back-react on the inflationary dynamics,
has been instead first discussed by Ratra~\cite{Ratra}, but it was ignored in the
seminal paper~\cite{Turner-Widrow} by Turner and Widrow on inflationary
magnetogenesis.

\subsection{IIIa. Strong coupling}

Let us consider the action for the electromagnetic field $A_\mu$,
\begin{equation}
\label{Action}
S_{\rm em} = \int \! d^4\!x \, \sqrt{-g} \, {\mathcal L}_{\rm em},
\end{equation}
where $g$ is the determinant of the metric tensor $g_{\mu\nu}$, and
${\mathcal L}_{\rm em}$ the electromagnetic Lagrangian density, and let us assume that
the electromagnetic field is coupled to the (homogeneous) inflaton field $\phi$
through a general coupling of the form
\begin{equation}
\label{Lagrangian1}
{\mathcal L}_{\rm em} = f(\phi){\mathcal L}_M + {\mathcal L}_{\rm int}.
\end{equation}
Here, $f(\phi)$ is a generic, positive-defined function of the inflaton,
${\mathcal L}_M =  -\frac14 F_{\mu\nu} F^{\mu\nu}$ is the standard free Maxwell
Lagrangian density, with $F_{\mu \nu} = \partial_\mu A_\nu - \partial_\nu A_\mu$,
while ${\mathcal L}_{\rm int} = j^\mu A_{\mu}$
is the standard interaction term with conserved external current.
If, for the sake of simplicity, we assume that $j_{\mu}$ is provided just by a charged
massless fermion fields $\psi$, we than have $j^{\mu} = e \bar{\psi} \gamma^\mu \psi$, where
$e$ is the electric charge, and $\gamma^\mu$ are the Dirac matrices in curved spacetime.
The latter are related to the standard Dirac matrices in Minkowski spacetime through
$\gamma^\mu = e^\mu_{\;\,a} \gamma^a$, with $e^{\mu}_{\;\,a}$ being the vierbein~\cite{Birrell-Davies}.
\footnote{The vierbein satisfies the condition $e^{\mu}_{\;\,a} e_{b\mu} = \eta_{ab}$, and
is such that $g_{\mu\nu} = e_\mu^{\;\,a} \, e_\nu^{\;\,b} \, \eta_{ab}$,
where $\eta_{ab}$ is the metric tensor in Minkowski spacetime.}
(In this paper, indices in Minkowski spacetime are indicated with the first letters of the Latin alphabet and run from 0 to 3.
Indices in curved spacetimes are indicated with Greek letters and run from 0 to 3.
Latin indices from the middle of the alphabet run from 1 to 3 and indicates spatial components
of a given tensor.)

Re-writing the Lagrangian density~(\ref{Lagrangian1}) as
\begin{equation}
\label{Lagrangianpsi}
{\mathcal L}_{\rm em} = f(\phi) \left( {\mathcal L}_M + \frac{e}{f(\phi)} \, \bar{\psi} \gamma^\mu \psi\right) \!,
\end{equation}
we see that the quantity $e/f(\phi)$ plays the role of an effective, time-dependent electric charge.
The case $f(\phi) \ll 1$ would then correspond to a strong coupling between the fermion and the electromagnetic fields,
and the theory would be in a (unmanageable) strong-coupling regime,
as firstly pointed out in~\cite{Demozzi}.
For this reason, we assume that $f(\phi) \gtrsim 1$ during inflation
and, obviously, $f(\phi) \simeq 1$ at the end of inflation in order to recover the standard electrodynamics.
Accordingly, we will consistently neglect, in the following, the interaction term ${\mathcal L}_{\rm int}$
in Eq.~(\ref{Lagrangian1}).

\subsection{IIIb. Backreaction}

It is tacitly assumed in the literature that inflationary magnetogenesis takes place in a fixed
curved spacetime background. Therefore, we must consistently check that
the vacuum expectation value (VEV) of the electromagnetic energy-momentum tensor
is always negligible with respect to the energy-momentum tensor of the inflaton.

The electromagnetic energy-momentum tensor can be found by varying the action
with respect to the metric tensor,
\begin{equation}
\label{Tdef}  (T_{\rm em})_{\mu\nu} = \frac{2}{\sqrt{-g}} \, \frac{\delta S_{\rm em}}{\delta g^{\mu\nu}} \, .
\end{equation}
We obtain $(T_{\rm em})_{\mu\nu} = f(\phi) (T_M)_{\mu\nu}$,
where $(T_M)_{\mu\nu} = F_{\alpha\mu} F_{\nu}^{\;\alpha} + \frac14 F_{\alpha\beta} F^{\alpha\beta} g_{\mu\nu}$
is the standard Maxwell energy-momentum tensor.

Let us restrict our analysis to the case of a spatially flat, Friedmann-Robertson-Walker universe,
described by the line element
\begin{equation}
\label{line}
ds^2 = a^2(d\eta^2 - d \x^2),
\end{equation}
where $\eta$ is the conformal time and $a(\eta)$ is the expansion parameter
[the latter is normalized to unity at the present conformal time $\eta_0$, $a(\eta_0) = 1$].
Moreover, we assume, for the sake of simplicity, that inflation is described by a de Sitter
phase. In this case, the conformal time is inversely proportional to the expansion parameter, $\eta = -1/Ha$,
and the Hubble parameter $H$ is a constant. Moreover, the energy-momentum
tensor of the inflaton is $(T^\mu_\nu)_{\rm inf} = M^4 \delta^\mu_\nu$, where
$\delta^\mu_\nu$ is the Kronecker delta. Here, $M$ is the scale of inflation,
related to the energy density of inflation, $\rho_{\rm inf}$, through
$M^4 = \rho_{\rm inf} = 3H^2/(8\pi G)$, where $G=1/\mPl^2$ is the Newton constant
and $\mPl \sim 10^{19} \GeV$ is the Planck mass.

Since both the background spacetime and the coupling function $f(\phi)$ are
homogeneous and isotropic, the VEV of the electromagnetic energy-momentum tensor
takes on the simple form
\begin{equation}
\label{Tdiag}
\langle (T_{\rm em})^\mu_\nu \rangle = \rho_{\rm em} \, \mbox{diag}(1,-1/3,-1/3,-1/3),
\end{equation}
where $\rho_{\rm em} = \langle(T_{\rm em})^0_0 \rangle$ is the VEV of the electromagnetic energy density.
(For the quantization of the theory and a formal definition of the vacuum, see section~V.)
Consequently, the condition that the electromagnetic backreaction on inflation is negligible can be expressed as
$\rho_{\rm em} \ll \rho_{\rm inf}$.
The electromagnetic energy density is made up of an
electric contribution and a magnetic part,
\begin{equation}
\label{rhoemphi}
\rho_{\rm em} = f(\phi) \! \left(\frac12 \langle \E^2 \rangle + \frac12 \langle \B^2 \rangle \right) \! ,
\end{equation}
where the electric and magnetic fields are defined as usual as
$a^2 \E = -\dot{\A}$ and $a^2 \B = \nabla \times \A$, with $A_\mu = (0,\A)$.
Here, and in the following, we work the Coulomb gauge, $A_0 = \partial_i A_i = 0$,
we denote the differentiation with respect to the conformal time with a dot,
and we use the symbol $\nabla$ for indicating the nabla operator in comoving coordinates.

The two-point correlators $\langle \E^2 \rangle$ and $\langle \B^2 \rangle$
are formally infinite due to the ultraviolet divergence
of the corresponding spectra. This kind of divergence, typical in quantum theory in curved spacetime,
can be cured by the standard techniques of renormalization, such as adiabatic renormalization.
Nevertheless, we are principally interested to large-scale electromagnetic modes which are outside
the horizon. These modes, which are expected to behave classically, belong
to the non-divergent, infrared part of the spectra.
Therefore, it is convenient to work in Fourier space and introduce the
the so-called electric and magnetic power spectra,
${\mathcal P}_{E}(k,\eta)$ and ${\mathcal P}_{B}(k,\eta)$, through
\begin{eqnarray}
\label{EBsquared}
\langle \E^2 \rangle = \int_0^{\infty} \! \frac{dk}{k} \, {\mathcal P}_{E}(k), \;\;\;
\langle \B^2 \rangle = \int_0^{\infty} \! \frac{dk}{k} \, {\mathcal P}_{B}(k).
\end{eqnarray}
The electromagnetic energy density stored on the mode $k$ is then
\begin{equation}
\label{rhoemphi}
\rho_{\rm em}(k) = f(\phi) \! \left[\frac12 {\mathcal P}_{E}(k) + \frac12 {\mathcal P}_{B}(k) \right] \! ,
\end{equation}
where $\rho_{\rm em}(k,\eta)$ is the electromagnetic energy spectrum defined by
$\rho_{\rm em} = \int_0^{\infty} \! \frac{dk}{k} \, \rho_{\rm em}(k)$.
The condition that the electromagnetic backreaction on inflation is negligible
can then be defined, mode-by-mode, by
\begin{equation}
\label{back}
\rho_{\rm em}(k) \lesssim \rho_{\rm inf} \sim 10^{10} \left(\frac{M}{10^{16}\GeV}\right)^{\!-4} H^4.
\end{equation}
A particularly interesting class of models is that for which the coupling function $f(\phi)$
scales in time as $f(\phi) \propto \eta^6$. This gives a scale-invariant magnetic spectrum,
to wit ${\mathcal P}_{B}(k)$ independent on $k$. The attractive figure of this model
resides in the fact that, as firstly pointed out in~\cite{Martin-Yokoyama},
all the existing constraints on cosmic magnetic fields do not
strongly peak over a specific range of either small or large scales. Hence, a
scaling-invariant magnetic field can satisfy, in a ``natural way'',
all the current experimental bounds, included the one in Eq.~(\ref{xiseed}).
For the scaling-invariant case, the electric and magnetic spectra are,
roughly speaking,
\begin{equation}
\label{E2B2}
{\mathcal P}_{E}(k) \sim \frac{{\mathcal P}_{B}(k)}{(-k\eta)^2} \, ,  \;\;\;
{\mathcal P}_{B}(k) \sim \frac{H^4}{f(\phi)} \, .
\end{equation}
On super-horizon scales ($-k\eta \ll 1$), then, the dominant contribution to the electromagnetic
energy-momentum tensor is provided by the electric part. Its maximum value is attained at the
end of inflation, $\eta = \eta_{\rm end}$. Therefore, backreaction on inflation is negligible on
scales $\lambda = 1/k$ such that
\begin{equation}
\label{lambda}
\lambda \lesssim \lambda_{\rm max} = 10^{-16} \left(\frac{M}{10^{16}\GeV}\right)^{\!\!-3} \Mpc,
\end{equation}
where we used the fact that
$-k\eta_{\rm end} \sim 10^{-22} (\Mpc/\lambda)(10^{16}\GeV/M)$.
To simplify the analysis, we have considered here
the case of instantaneous reheating, to wit, we have assumed that after inflation the
Universe entered directly in the radiation dominated era.
Needless to say, the electromagnetic backreaction on inflation
has to be negligible on all observable scales. This, in turns, means that
$\lambda_{\rm max}$ has to be greater than the present horizon scale, $H_0^{-1} \simeq 4000 \Mpc$,
where $H_0$ is the Hubble constant. Accordingly, the scale of inflation has to be below $10^9 \GeV$.
Such a low scale seems to be incompatible with recent results
on the detection of inflation-produced gravitational waves, which require a scale of
inflation around $10^{16}\GeV$~\cite{BICEP2}.
However, even assuming a scale as low as $M \sim 10^9 \GeV$, the amplitude of the
inflation-produced magnetic field would be today too small to directly explain cosmic magnetism.
Indeed, the actual magnetic field for the scaling-invariant case is
\begin{equation}
\label{B0intro} B_0 \sim 10^{-12} \left(\frac{M}{10^{16}\GeV}\right)^{\!\!2} \G,
\end{equation}
and it assumes the extremely low value $B_0 \sim 10^{-26} \G$ for $M \sim 10^9 \GeV$.

\section{IV. Lorentz-violating couplings}

The arguments in section III clearly show that the generation of (scaling-invariant)
magnetic fields during (de Sitter) inflation, able to directly explain cosmic magnetization,
is problematic due to their strong backreaction effects.
The validity of this sort of no-go theorem for inflationary magnetogenesis, however,
is not general, but it is restricted to the specific model described by the Lagrangian density~(\ref{Lagrangian1}).
This leaves open the possibility to explore different couplings between the inflaton
and the photon that may eventually generate, in a self-consistent way,
cosmic magnetic fields.

In the following, we investigate one such a possibility, by looking at
a possible new interaction of the inflaton with the electromagnetic field,
this time in the context of the Lorentz-violating
extension of the standard model of particle physics (for other mechanisms of cosmic magnetic fields
at inflation reposing on the violation of Lorentz symmetry
see~\cite{Generation2a,Generation2b,Generation2c,Generation2d,Generation2e,Generation2f,Generation2g,Generation2h,Generation2i}).

\subsection{IVa. Lagrangian}

The photon sector of the minimal Lorentz-violating standard model extension (SME) is described by
the action
\begin{equation}
\label{actionLV}
S_\gamma = \int \! d^4\!x \, \sqrt{-g} \, ({\mathcal L}_M + {\mathcal L}_{LV}),
\end{equation}
where ${\mathcal L}_M$ is the Lorentz- and CPT-invariant Maxwell Lagrangian density, while
\begin{equation}
\label{Lagrangiangamma1}
{\mathcal L}_{LV} = {\mathcal L}_{\rm CPT-even} + {\mathcal L}_{\rm CPT-odd}
\end{equation}
contains all Lorentz-violating terms that involve the photon field. They can be separated
into two parts, ${\mathcal L}_{\rm CPT-even}$ and ${\mathcal L}_{\rm CPT-odd}$, with
the former preserving and the latter violating CPT symmetry, respectively.

In Minkowski spacetime, Lorentz violation is achieved
by coupling the electromagnetic field to rank-$n$, constant spacetime tensors $k_{a_1a_2...a_n}$,
known as external or background tensors.
The passage from Minkowski to a general curved spacetime is obtained
via the vierbein $e_{\mu}^{\;\,a}$,
$k_{\mu_1\mu_2...\mu_n} = e_{\mu_1}^{\;\,a_1} e_{\mu_2}^{\;\,a_2} ... e_{\mu_n}^{\;\,a_n} k_{a_1a_2...a_n}$~\cite{Kostelecky2}.
In this passage, however, the external tensors acquire a spacetime dependence
and then cease to be constant. This is due to the fact that
the vierbein $e_{\mu}^{\;\,a}(x)$ is, generally, a function of the spacetime position $x$.

In the photon sector, the most general renormalizable Lagrangian density
contains only three Lorentz-violating terms,
\begin{eqnarray}
\label{Lagrangiangamma2}
&& {\mathcal L}_{\rm CPT-even} = - \frac14 (k_F)_{\mu\nu\alpha\beta} F^{\mu\nu} F^{\alpha\beta},
\\
\label{Lagrangiangamma3}
&& {\mathcal L}_{\rm CPT-odd} = \frac12 (k_{AF})_{\mu} A_\nu \widetilde{F}^{\mu\nu} - (k_A)_{\mu} A^\mu,
\end{eqnarray}
where $(k_F)_{\mu\nu\alpha\beta}$, $(k_{AF})_{\mu}$, and $(k_A)_{\mu}$,
are background tensors. Their components are arbitrary real spacetime functions and are
known as coefficients for Lorentz violation. Although the presence of the
external tensors may indicate an explicit breaking of Lorentz violation,
the form of the Lagrangian terms~(\ref{Lagrangiangamma2})-(\ref{Lagrangiangamma3})
is completely general and independent of the origin of the Lorentz violation.
Indeed, these terms would have the same form in the case where Lorentz violation
were spontaneous, deriving, for example, from the fact that the external tensor
$k_{\mu_1\mu_2...\mu_n}$ are vacuum expectation values of corresponding field operators
$\mathcal{K}_{\mu_1\mu_2...\mu_n}$,
\begin{equation}
\label{vacK}
k_{\mu_1\mu_2...\mu_n} = \langle 0| \mathcal{K}_{\mu_1\mu_2...\mu_n} |0\rangle.
\end{equation}
We now generalize the coupling in Eq.~(\ref{Lagrangian1}) by assuming that
the inflaton is coupled to the photon field, via the generic function $f(\phi)$,
to both the standard photon kinetic term ${\mathcal L}_M$,
and the Lorentz-violating term ${\mathcal L}_{LV}$,
\begin{equation}
\label{noname}
{\mathcal L}_{\rm em} = f(\phi)({\mathcal L}_M + {\mathcal L}_{LV}).
\end{equation}
For the sake of simplicity, we consider only the CPT-even terms in the Lagrangian density~(\ref{Lagrangiangamma1}).
The electromagnetic Lagrangian density then reads
\begin{equation}
\label{Lagrangian2bis}
{\mathcal L}_{\rm em} = f(\phi){\mathcal L}_{MK},
\end{equation}
where
\begin{equation}
\label{LagrangianMK}
{\mathcal L}_{MK} = {\mathcal L}_M - \frac14 (k_F)_{\mu\nu\alpha\beta} F^{\mu\nu} F^{\alpha\beta}
\end{equation}
is referred to as the Maxwell-Kosteleck\'{y} Lagrangian density.
The dimensionless rank-4 background tensor $(k_F)_{\mu\nu\alpha\beta}$
is antisymmetric on the first two and last two indices, and it is symmetric
for the interchange of the first and last pair of indices. These symmetries reduce
the number of independent components of $(k_F)_{\mu\nu\alpha\beta}$ to 21.
It is useful to decompose $(k_F)_{\mu\nu\alpha\beta}$ into irreducible multiplets~\cite{Kostelecky2},
$21 = 1_a + 1_s + 9_s + 10_s$, where $1_a$ represents an antisymmetric singlet (pseudoscalar),
$1_s$ a symmetric singlet (scalar), $9_s$ a symmetric traceless rank-2 tensor, and
$10_s$ a rank-4 tensor possessing the same symmetries of $(k_F)_{\mu\nu\alpha\beta}$
and such that any contraction is identically zero.

Let us restrict our analysis to the case where the background tensor in Eq.~(\ref{LagrangianMK})
is constructed from fundamental (not composite) tensors which appear just once in the definition of
$(k_F)_{\mu\nu\alpha\beta}$. Excluding the cases where such fundamental tensors are a scalar
and/or a pseudoscalar, in which case the resulting theory is Lorentz invariant,
we are left with the cases of a fundamental rank-2 symmetric tensor and/or a fundamental rank-4 tensor.
In this paper, and for the sake of simplicity, we consider just the case of a
fundamental rank-2 tensor $(k_F)_{\mu\nu}$, and leave the case of a rank-4 tensor to
future investigations.
In this specific case, the independent components of $(k_F)_{\mu\nu\alpha\beta}$ reduce to 10,
and they are given by $k_F = (k_F)^\mu_\mu$ and $(\widehat{k}_F)_{\mu\nu} = (k_F)_{\mu\nu} - \frac14 k_F g_{\mu\nu}$,
which are, respectively, the trace and the traceless part of the tensor $(k_F)_{\mu\nu}$.
Accordingly, the electromagnetic Lorentz-violating Lagrangian density,
which describe the coupling between the inflaton and the photon, can be written as
\begin{equation}
\label{Lagrangian2}
{\mathcal L}_{\rm em} = f(\phi) \! \left( \! {\mathcal L}_M
+ \frac14 \xi_1 k_F F_{\alpha\beta} F^{\alpha\beta}
- \xi_2 (\widehat{k}_F)_{\mu\nu} F^{\mu\alpha} F^\nu_{\;\;\, \alpha} \! \right) \! ,
\end{equation}
where $\xi_i$ are real numerical factors. The background tensor $(k_F)_{\mu\nu\alpha\beta}$,
when expressed as a function of the fundamental rank-2 tensor $(k_F)_{\mu\nu}$, has the form
\begin{equation}
\label{kFdec}
(k_F)_{\mu\nu\alpha\beta} = -\xi_1 k_F g_{\mu[\alpha} g_{\beta]\nu} +
4\xi_2 (\widehat{k}_F)_{\mu][\alpha} g_{\beta][\nu} \, ,
\end{equation}
where square brackets $[...]$ indicate antisymmetrization
of the indices enclosed, e.g., $T_{\mu_1 ... [\mu_i \mu_j] ... \mu_n} =
\frac12 (T_{\mu_1 ...\mu_i \mu_j... \mu_n} - T_{\mu_1 ... \mu_j \mu_i ... \mu_n})$
and $T_{\mu_1] ... [\mu_n} = \frac12 (T_{\mu_1 ... \mu_n} - T_{\mu_n ... \mu_1})$.

\subsection{IVb. Equation of motion}

As in section III, we restrict our analysis to the case of a spatially flat,
Friedmann-Robertson-Walker universe.
Since $g_{\mu\nu} = a^2 \eta_{\mu\nu}$, we can take for the vierbein
$e_\mu^{\;\,b} = a\delta_\mu^b$, 
where $\delta_\mu^b$ is the Kronecker delta.
Accordingly, we have
$(k_F)_{\mu\nu} = e_\mu^{\;\,b} e_\nu^{\;\,c} (k_F)_{cb} = a^2 \delta_\mu^b \delta_\nu^c (k_F)_{cb}$.
Let now assume that the background tensor $(k_F)_{ab}$ is homogeneous and isotropic,
so that the number of its independent components reduces to 2. In this case, $(k_F)_{ab}$
can be generally written as $(k_F)_{cb} = \mbox{diag}(\rho_K,p_K,p_K,p_K)$,
where $\rho_K$ and $p_K$ are two scalar functions which depend only on the
conformal time $\eta$. In curved spacetime, then,
the background tensor assumes the form
\begin{equation}
\label{kmunu}
(k_F)^\mu_\nu = \mbox{diag}(\rho_K,-p_K,-p_K,-p_K).
\end{equation}
It is useful, for the following discussion, to introduce the electromagnetic Lagrangian $L_{\rm em}$
through
\begin{equation}
\label{Actionbis}
S_{\rm em} = \int \! d\eta\, L_{\rm em}.
\end{equation}
Taking into account Eq.~(\ref{Action}) and the fact that $\sqrt{-g} = a^4$, we have
\begin{equation}
\label{LLL}
L_{\rm em} = \int d^3\!x \, a^4 {\mathcal L}_{\rm em},
\end{equation}
with ${\mathcal L}_{\rm em}$ given by Eq.~(\ref{Lagrangian2}).
Working in the Coulomb gauge, we have
\begin{equation}
\label{LLLbis}
L_{\rm em} = \int \! d^3\!x \left( \frac12 \, \varepsilon \dot{\A}^2 - \frac{1}{2\mu} (\nabla \A)^2 \right) \!,
\end{equation}
where we have defined the time-dependent functions
\begin{eqnarray}
\label{g1}
&& \!\!\!\!\!\!\!\!\!\!\!\!\!\!\! \varepsilon =
f(\phi) [1 - (\xi_1 - \xi_2 ) \rho_K + (3\xi_1 + \xi_2) p_K], \\
\label{g2}
&& \!\!\!\!\!\!\!\!\!\!\!\!\!\!\! \mu^{-1} =
f(\phi) [1 - (\xi_1 + \xi_2) \rho_K + (3\xi_1 - \xi_2) p_K].
\end{eqnarray}
Varying the action~(\ref{LLLbis}) with respect to $\A$, we find the equation of motion
for the vector potential,
\begin{equation}
\label{vecEqmotion}
\ddot{\A} + \frac{\dot{\varepsilon}}{\varepsilon} \, \dot{\A} - \frac{1}{n^2} \nabla^2 \A = 0,
\end{equation}
where we have defined $n = \sqrt{\varepsilon \mu}$,
and we assume that $\varepsilon$ and $\mu$ are positive-defined quantities.

\subsection{IVc. Analogy with continuous media}

It is well known in the literature that there exists an analogy between the photon sector of the minimal SME
and the electrodynamics of continuous (or macroscopic) media~\cite{Colladay}.
In our case, this analogy works as follows.
We rewrite the electromagnetic Lagrangian density as
\begin{equation}
\label{Eom1}
{\mathcal L}_{{\rm em}} = {\mathcal L}_M
+ \frac14 \, \chi^{\mu \nu}_{\;\;\;\, \alpha \beta} F_{\mu \nu} F^{\alpha \beta}
\end{equation}
where
\begin{eqnarray}
\label{Eom2}
\chi^{\mu \nu}_{\;\;\;\, \alpha \beta} \!\!& = &\!\!
\frac12 \, \{ 1 - f(\phi) [1 - (\xi_1 + \xi_2) k_F] \} \delta^{\mu\nu}_{\alpha\beta} \nonumber \\
\!\!& - &\!\! 4 \xi_2 f(\phi) \, \delta^{[\mu}_{[\alpha} (k_F)^{\nu]}_{\beta]},
\end{eqnarray}
is the susceptibility tensor, and $\delta^{\mu\nu}_{\alpha\beta}$ the generalized Kronecker delta.
Introducing the polarization-magnetization tensor $\mathcal{M}^{\mu \nu}$ as
\begin{equation}
\label{Eom3}
\mathcal{M}^{\mu \nu} = \chi^{\mu \nu}_{\;\;\;\, \alpha \beta} F^{\alpha \beta},
\end{equation}
the equation of motion is
\begin{equation}
\label{Eom4}
\mathcal{D}^{\mu \nu}_{\;\;\;\; ;\mu} = 0,
\end{equation}
where
\begin{equation}
\label{Eom5}
\mathcal{D}^{\mu \nu} = F^{\mu \nu} - \mathcal{M}^{\mu \nu}
\end{equation}
is the displacement tensor.
In the Coulomb gauge, Eq.~(\ref{Eom4}) reduces to $0=0$ for $\nu = 0$,
and to Eq.~(\ref{vecEqmotion}) for $\nu = i$.

Let us introduce the electric and magnetic fields,
$a^2 E_i = -F_{0i}$ and $a^2 B_i = \frac12 \, \epsilon_{ijk} F_{jk}$,
the displacement and magnetizing fields,
$a^2 D_i = -\mathcal{D}_{0i}$ and $a^2 H_i = \frac12 \, \epsilon_{ijk} \mathcal{D}_{jk}$,
and the polarization and magnetization fields,
$a^2 P_i = \mathcal{M}_{0i}$ and $a^2 M_i = \frac12 \, \epsilon_{ijk} \mathcal{M}_{jk}$.
Equation~(\ref{Eom5}) can then be rewritten, in three-dimensional form, as
$\D = \E + \PP$ and $\HH = \B - \M$,
where $\X = (X_1,X_2,X_3)$, and $\X$ stands for $\E$, $\B$, $\D$, $\HH$, $\PP$, or $\M$.
Equation~(\ref{Eom3}) gives, instead,
$\PP = (\varepsilon - 1) \E$ and $\M  =  (1 - \mu^{-1}) \B$,
so that
\begin{eqnarray}
\label{Eom9}
&& \D  = \varepsilon \E, \\
\label{Eom10}
&& \HH = \mu^{-1} \B.
\end{eqnarray}
The equations connecting the displacement and magnetizing fields
to the electric and magnetic fields are known, in the electrodynamic theory of continuous media,
as ``constitutive relations'', and completely determine (together with the boundary conditions)
the propagation properties of electromagnetic signals.
In particular, Eqs.~(\ref{Eom9}) and (\ref{Eom10}) describe an isotropic linear medium with
electric permittivity $\varepsilon$ and magnetic permeability $\mu$.
Accordingly, the evolution in vacuum of electromagnetic fields described by the
Lorentz-violating electromagnetic Lagrangian density~(\ref{Eom1}),
is formally equivalent to the evolution of electromagnetic fields described by the standard Maxwell theory
in a continuous medium with $\varepsilon$ and $\mu$ given by Eqs.~(\ref{g1}) and (\ref{g2}).
Continuing with the analogy of continuous media,
the quantity $n$ defined below Eq.~(\ref{vecEqmotion}) can be interpreted as the refractive index
of the medium.

Finally, and for the sake of completeness, we observe that the equation of motion,
in terms of the displacement and magnetizing fields, assume the form
\begin{equation}
\label{EqDH}
\nabla \cdot (a^2 \D) = 0, \;\;\; \frac{\partial (a^2\D)}{\partial \eta} = \nabla \times (a^2\HH),
\end{equation}
while the Bianchi identities are
\begin{equation}
\label{Bianchi}
~~~~~~\, \nabla \cdot (a^2\B) = 0, \;\;\; \frac{\partial (a^2\B)}{\partial \eta} = -\nabla \times (a^2\E).
\end{equation}
Inserting Eqs.~(\ref{Eom9}) and (\ref{Eom10}) in the second equation of Eq.~(\ref{EqDH}),
we recover Eq.~(\ref{vecEqmotion}).

\section{V. Quantization}

Let us now quantize the electromagnetic field whose dynamics is described by Lagrangian~(\ref{LLLbis}).

\subsection{Va. Wronskian condition}

We expand the electromagnetic vector potential as
\begin{equation}
\label{A2}
{\A}(\eta,\x) = \sum_{\lambda=1}^2 \int \!\! \frac{d^3k}{(2\pi)^3 \sqrt{2k}} \, \ee_{\kk,\lambda} \,
a_{\kk,\lambda} \, A_{k,\lambda}(\eta) \, e^{i\kk \x} + \mbox{H.c.} ,
\end{equation}
where $\kk$ is the comoving wavenumber, with $k = |\kk|$, and
$\ee_{\kk,\lambda}$ are the standard circular polarization vectors.
\footnote{The vectors $\ee_{\kk,\lambda}$ satisfy the following properties:
($i$) $\kk \cdot \ee_{\kk,\lambda} = 0$,
($ii$) $\ee_{\kk,\lambda} \cdot \ee_{\kk,\lambda'}^* = \delta_{\lambda \lambda'}$,
($iii$) $\sum_{\lambda} (\ee_{\kk,\lambda})_i (\ee_{\kk,\lambda'}^*)_j = \delta_{ij} - \hat{k}_i \hat{k}_j$,
($iv$) $\ee_{-\kk,\lambda}^* = -\ee_{\kk,\lambda}$, and
($v$) $i \hat{\kk} \times \ee_{\kk,\lambda} = (-1)^{\lambda+1} \ee_{\kk,\lambda}$, where $\hat{\kk} = \kk/k$.}
The annihilation and creation operators $a_{\kk,\lambda}$ and $a_{\kk,\lambda}^{\dag}$
satisfy the usual commutation relations
\begin{eqnarray}
\label{x1}
&& [a_{\kk,\lambda}, a_{\kk',\lambda'}^{\dag}] = (2\pi)^3 \delta_{\lambda \lambda'} \delta(\kk-\kk'), \\
\label{x2}
&& [a_{\kk,\lambda}, a_{\kk',\lambda'}] = [a_{\kk,\lambda}^{\dag}, a_{\kk',\lambda'}^{\dag}] = 0.
\end{eqnarray}
The vacuum state $|0\rangle$
is defined by $a_{\kk,\lambda} |0\rangle = 0$ for all $\kk$ and $\lambda$, and it is normalized as
$\langle 0|0\rangle = 1$.

The equation of motion for the two photon polarization states, $A_{k,\lambda}$, is obtained by inserting
Eq.~(\ref{A2}) in Eq.~(\ref{vecEqmotion}),
\begin{equation}
\label{motion1}
\ddot{A}_{k,\lambda} + \frac{\dot{\varepsilon}}{\varepsilon} \, \dot{A}_{k,\lambda} + \frac{k^2}{n^2} \, A_{k,\lambda} = 0.
\end{equation}
In order to have a consistent quantization of the electromagnetic field, the solutions of the above equation
must satisfy a normalization condition, known as the Wronskian condition, which can be obtained as follows.
Let us introduce the electromagnetic conjugate momentum, $\boldsymbol{\pi} = (\pi^1,\pi^2,\pi^3)$,
as usual as
\begin{equation}
\label{comm1}
\boldsymbol{\pi} = \frac{\delta L_{\rm em}}{\delta \dot{\A}} = \varepsilon \dot{\A},
\end{equation}
where in the last equality we used Eq.~(\ref{LLLbis}), and let us impose the canonical commutation relation
\begin{equation}
\label{comm2}
[A_i(\x),\pi^j(\y)] = i \delta_{\perp \, i}^j(\x-\y),
\end{equation}
where
\begin{equation}
\label{comm3}
\delta^{\perp}_{ij} = \int \!\! \frac{d^3 k}{(2\pi)^3} \, e^{i \kk (\x-\y)} (\delta_{ij} - \hat{k}_i \hat{k}_j)
\end{equation}
is the transverse delta function.
Inserting Eqs.~(\ref{A2}) and (\ref{comm1}) in the left hand side of Eq.~(\ref{comm2}),
we find that the latter equation is satisfied only if
\begin{equation}
\label{comm4}
\sum_{\lambda=1}^2 \ee_{\kk,\lambda} \otimes \ee_{\kk,\lambda}^* \!
\left( \! W[A_{k,\lambda},A_{k,\lambda}^*] - \frac{2ik}{\varepsilon} \right ) = 0,
\end{equation}
where we used Eqs.~(\ref{x1})-(\ref{x2}). Here,
\begin{equation}
\label{Wdef}
W[A_{k,\lambda}^{(1)},A_{k,\lambda}^{(2)}]
= A_{k,\lambda}^{(1)} \dot{A}_{k,\lambda}^{(2)} - \dot{A}_{k,\lambda}^{(1)} A_{k,\lambda}^{(2)}
\end{equation}
is the Wronskian of any two independent solutions, $A_{k,\lambda}^{(1)}(\eta)$ and $A_{k,\lambda}^{(2)}(\eta)$,
of Eq.~(\ref{motion1}).
Using the Abel identity~\cite{Abel}, the above Wronskian can be found explicitly,
\begin{equation}
\label{Wbis}
W[A_{k,\lambda}^{(1)}(\eta),A_{k,\lambda}^{(2)}(\eta)] =
W[A_{k,\lambda}^{(1)}(\eta_i),A_{k,\lambda}^{(2)}(\eta_i)] \, \frac{\varepsilon(\eta_i)}{\varepsilon(\eta)} \, ,
\end{equation}
where $\eta_i$ is an arbitrary time.
Accordingly, Eq.~(\ref{comm4}) can be written as
\begin{equation}
\label{comm5}
\sum_{\lambda=1}^2 \ee_{\kk,\lambda} \otimes \ee_{\kk,\lambda}^* \!
\left( \! W[A_{k,\lambda}(\eta_i),A_{k,\lambda}^*(\eta_i)] -\frac{2ik}{\varepsilon(\eta_i)} \right ) = 0.
\end{equation}
We now take $\eta_i$ as the initial time, namely when inflation begins,
and we assume that $A_{k,1}(\eta_i) = A_{k,2}(\eta_i)$.
This choice implies that $W[A_{k,\lambda}(\eta_i),A_{k,\lambda}^*(\eta_i)]$
does not depend on $\lambda$. Consequently, Eq.~(\ref{comm5}) is satisfied only if
\begin{equation}
\label{comm7}
W[A_{k,\lambda}(\eta_i),A_{k,\lambda}^*(\eta_i)] = \frac{2ik}{\varepsilon(\eta_i)} \, ,
\end{equation}
in which case we have
\begin{equation}
\label{comm8}
W[A_{k,\lambda}(\eta),A_{k,\lambda}^*(\eta)] = \frac{2ik}{\varepsilon(\eta)},
\end{equation}
for all $\eta$. Equation~(\ref{comm8}) represents the desired condition that must be satisfied
by any solution $A_{k,\lambda}(\eta)$ of the equation of motion in order to have a consistent
quantization of the electromagnetic field.

\subsection{Vb. Bunch-Davies normalized solutions}

Let us assume, for the sake of simplicity, that the external tensor $(k_F)_\nu^\mu$
is constant during inflation
\footnote{It is important to stress that $(k_F)_{\mu \nu}$ must evolve after inflation
in such a way to be consistent with current experimental limits on Lorentz-violation coefficients~\cite{Data}.}
(so that $\rho_k$ and $p_K$ are constant as well),
and that the coupling function $f(\phi)$ evolves in time following a simple power law,
\begin{equation}
\label{geta}
f(\phi(\eta)) = f_i \! \left(\frac{\eta}{\eta_i}\right)^{\! \gamma} \! ,
\end{equation}
where $f_i = f(\phi(\eta_i))$, and $\gamma$ is a free index.
In this case, the permittivity $\varepsilon$ and the permeability $\mu$ evolve in time
as $\eta^\gamma$, while the refractive index $n$ is a constant.
The solution of Eq.~(\ref{motion1}) is then easily found,
\begin{equation}
\label{sol1}
A_{k,\lambda} = (-k\eta)^\nu \! \left[ c_k^{(1)} H_\nu^{(1)}(-k\eta/n) + c_k^{(2)} H_\nu^{(2)}(-k\eta/n) \right] \!,
\end{equation}
where $\nu = (1-\gamma)/2$, $H_\nu^{(1,2)}(x)$ are the Hankel functions of first and second kind, respectively,
and $c_k^{(1,2)}$ are integration constants. The latter can be fixed by the choice of the vacuum,
which we take to be the Bunch-Davies vacuum~\cite{Birrell-Davies,Parker-Toms}. It reduces to the standard Minkowski vacuum
in the short wavelength limit, $k \rightarrow \infty$. To find it,
let us re-scale the electromagnetic field as
\begin{equation}
\label{psi-A1}
\psi_{k,\lambda} = \frac{A_{k,\lambda}}{\sqrt{|W[A_{k,\lambda},A_{k,\lambda}^*]|}} \, .
\end{equation}
Inserting Eq.~(\ref{psi-A1}) in Eq.~(\ref{motion1}), we see that
the re-scaled $\psi$-modes satisfy the equation of motion
\begin{equation}
\label{Sch}
\ddot{\psi}_{k,\lambda} = U_{k} {\psi}_{k,\lambda},
\end{equation}
where we have defined
\begin{equation}
\label{Uk}
U_{k} = - \frac{k^2}{n^2} + \frac{1}{\sqrt{\varepsilon}} \,\frac{\partial^2}{\partial \eta^2} \sqrt{\varepsilon} \, .
\end{equation}
Let us observe that Eq.~(\ref{Sch}) is formally equal to the zero-mode, one-dimensional Schrodinger equation
with potential energy $U_{k}$, $\eta$ taking the place of the spatial coordinate,
and $k$ playing the role of a free constant parameter.
If $\psi_{k,\lambda}^{(1)}$ and $\psi_{k,\lambda}^{(2)}$ are any two solutions of Eq.~(\ref{Sch}),
the following inner product is conserved,
\begin{equation}
\label{inner}
\langle \psi_{k,\lambda}^{(1)} | \psi_{k,\lambda}^{(2)} \rangle =
-i \! \left(\psi_{k,\lambda}^{(1)} \dot{\psi}_{k,\lambda}^{(2)} - \dot{\psi}_{k,\lambda}^{(1)} \psi_{k,\lambda}^{(2)}\right) \! .
\end{equation}
Moreover, using Eq.~(\ref{comm8}), we see that $\psi$-modes are normalized as
\begin{equation}
\label{inner2} \langle \psi_{k,\lambda} | \psi_{k,\lambda}^{*} \rangle = 1.
\end{equation}
For $k \rightarrow \infty$, the potential energy
is dominated by the first term in the right-hand-side of Eq.~(\ref{Uk}). Therefore,
the positive-frequency solution of Eq.~(\ref{Sch}) in the short wavelength limit is
$\psi_{k,\lambda} = c_k \, e^{-i\omega_k\eta}$, where $\omega_k = k/n$
and $c_k$ is an integration constant. The latter is fixed
the normalization condition~(\ref{inner2}),
$c_k = 1/\sqrt{2\omega_k}$, so that
$\psi_{k,\lambda} = 1/\sqrt{2\omega_k} \, e^{-i\omega_k\eta}$.
Accordingly, the Minkowski vacuum ($k \rightarrow \infty$)
is defined by the normalized electromagnetic field solution
\begin{equation}
\label{p2} A_{k,\lambda} = \sqrt{Z} \, e^{-i\omega_k\eta},
\end{equation}
where $Z = \sqrt{\mu/\varepsilon}$ is, in the language
of the electrodynamics of continuous media,
the wave impedance of the medium.

Equation~(\ref{sol1}) must then reduce to Eq.~(\ref{p2}) in the limit $k \rightarrow \infty$.
This happens only if
$c_k^{(1)} = [\pi e^{i\pi (\nu+1/2)}(-k\eta_i)^\gamma/2\varepsilon(\eta_i)]^{1/2}$ and $c_k^{(2)} = 0$,
in which case
\begin{equation}
\label{p3} A_{k,\lambda} =
\sqrt{\frac{\pi}{2}} \, e^{i\frac{\pi}{2}(\nu+1/2)} \sqrt{Z} \sqrt{-\omega_k\eta} \, H_\nu^{(1)}(-\omega_k\eta)
\end{equation}
is the desired Bunch-Davies vacuum normalized solution.

\section{VI. Backreaction on inflation}

We now draw our attention to the electromagnetic backreaction on inflation in the model
described by Lagrangian~(\ref{Lagrangian2}). We will find the conditions
under which such a backreaction is completely negligible.

Let us first observe that Lagrangian density~(\ref{Eom1}) can be conveniently rewritten as
\begin{equation}
\label{LLLtris} \mathcal{L}_{\rm em} = -\frac14 F_{\mu\nu} \mathcal{D}^{\mu\nu},
\end{equation}
with $\mathcal{D}_{\mu\nu}$ given by Eq.~(\ref{Eom5}).
The electromagnetic energy-momentum tensor is obtained by inserting  Eq.~(\ref{LLLtris}) in Eq.~(\ref{Tdef}).
We find
\begin{equation}
\label{TTT1}  (T_{\rm em})_{\mu\nu} = (T_{\rm med})_{\mu\nu} + (T_{\rm \mathcal{X}})_{\mu\nu}.
\end{equation}
Here,
\begin{equation}
\label{TTT2}  (T_{\rm med})_{\mu\nu} = F_{\alpha\{\mu} \mathcal{D}_{\nu\}}^{\;\;\alpha}
+ \frac14 F_{\alpha\beta} \mathcal{D}^{\alpha\beta} g_{\mu\nu}
\end{equation}
is the standard electromagnetic energy-momentum tensor in a medium described by
the displacement tensor~(\ref{Eom5}) [curly brackets $\{...\}$ indicate a symmetrization
of the indices enclosed, e.g., $T_{\mu_1 ... \{\mu_i \mu_j\} ... \mu_n} =
\frac12 (T_{\mu_1 ...\mu_i \mu_j... \mu_n} + T_{\mu_1 ... \mu_j \mu_i ... \mu_n})$],
and
\begin{equation}
\label{TTT3} (T_{\rm \mathcal{X}})_{\mu\nu} =
\frac14 \, \mathcal{X}^{\alpha\beta}_{~~~\gamma\delta\mu\nu} \, F_{\alpha\beta} F^{\gamma \delta}.
\end{equation}
The rank-five tensor
\begin{equation}
\label{TTT4}  \mathcal{X}^{\alpha\beta}_{~~~\gamma\delta\mu\nu} = \frac{2}{\sqrt{-g}}
\frac{\delta}{\delta g^{\mu\nu}} \! \int \! d^4\!x \, \sqrt{-g} \, \chi^{\alpha\beta}_{~~~\gamma \delta}
\end{equation}
is antisymmetric on the first two and second two indices, symmetric in the last two indices,
and it is symmetric for the interchange of the first and second pair of indices.

When the susceptibility tensor has the form of Eq.~(\ref{Eom2}), the electromagnetic energy-momentum tensor
can be written, in its full form, as
\begin{eqnarray}
\label{T} (T_{\rm em})_{\mu\nu} \!\!& = &\!\!
f(\phi) \! \left[ F_{\alpha\mu} F_{\nu}^{\;\alpha} + \frac14 F_{\alpha\beta} F^{\alpha\beta} g_{\mu\nu} \right. \nonumber \\
\!\!& - &\!\!  \left. \frac12 (\xi_1 + \xi_2) (k_F)_{\mu\nu} F_{\alpha\beta} F^{\alpha\beta} \right. \nonumber \\
\!\!& - &\!\! \left. 2 \xi_2 (k_F)^{\alpha}_{\beta} F_{\alpha\mu} F_{\nu}^{\;\, \beta}
+ 4 \xi_2 (k_F)^{\alpha}_{\{\mu} F_{\nu\}\beta} F_{\alpha}^{\;\;\beta} \right. \nonumber \\
\!\!& - &\!\! \left. \xi_2 (k_F)^{\alpha}_{\beta} F_{\alpha\gamma} F^{\beta\gamma} g_{\mu\nu} \right] \! .
\end{eqnarray}
Due to symmetry, we find that the only (possible) non-null components of the vacuum expectation value
of the electromagnetic energy-momentum tensor, $\langle (T_{\rm em})^\mu_\nu \rangle$, are
the electromagnetic energy density, $\langle (T_{\rm em})^0_0 \rangle$, and $\langle (T_{\rm em})^i_j \rangle =
\frac13 \left(\langle (T_{\rm em})^\mu_\mu \rangle - \langle (T_{\rm em})^0_0 \rangle \right) \! \delta_{ij}$.
Here, $(T_{\rm em})^\mu_\mu$ is the trace of the electromagnetic energy-momentum tensor,
which is, in general, different from zero due to the coupling of the photon to the
background tensor $(k_F)^\mu_\nu$. In particular, we have
\begin{eqnarray}
\label{TS1}
&& \rho_{\rm em} =
\frac{\varepsilon}{2} \, \langle \E^2 \rangle + \frac{1}{2\mu} \, \langle \B^2 \rangle
+ \langle (T_{\rm \mathcal{X}})^0_0 \rangle, \\
\label{TS2}
&& T_{\rm em} = \langle (T_{\rm \mathcal{X}})^\mu_\mu \rangle,
\end{eqnarray}
where we have defined
\begin{equation}
\label{rhoTdef}
\rho_{\rm em} = \langle (T_{\rm em})^0_0 \rangle, \;\;\; T_{\rm em} = \langle (T_{\rm em})^\mu_\mu \rangle.
\end{equation}
When the susceptibility tensor has the form in Eq.~(\ref{Eom2}), we have
\begin{eqnarray}
\label{TX0}
\!\!\!\!\!\!\!\!\!\!\!\!\!\!\!\!\!\! && \langle (T_{\rm \mathcal{X}})^0_0 \rangle =
f(\phi) \rho_K \! \left[ -(\xi_1 - \xi_2) \langle \E^2 \rangle + (\xi_1 + \xi_2) \langle \B^2 \rangle \right] \! , \\
\!\!\!\!\!\!\!\!\!\!\!\!\!\!\!\!\!\! && \langle (T_{\rm \mathcal{X}})^i_i \rangle =
f(\phi) p_K \! \left[ (3\xi_1 + \xi_2) \langle \E^2 \rangle - (3\xi_1 - \xi_2) \langle \B^2 \rangle \right] \! .
\end{eqnarray}
The vacuum expectation value of the squared magnetic and electric fields operator are easily found:
\begin{eqnarray}
\label{Bsquared} && \langle \B^2 \rangle = \int_0^{\infty} \! \frac{dk}{k} \, {\mathcal P}_{B}(k,\eta), \\
\label{Esquared} && \langle \E^2 \rangle = \int_0^{\infty} \! \frac{dk}{k} \, {\mathcal P}_{E}(k,\eta),
\end{eqnarray}
where ${\mathcal P}_{B}(k,\eta)$ and ${\mathcal P}_{E}(k,\eta)$ are the so-called magnetic and electric power spectra,
\begin{eqnarray}
\label{PB} && {\mathcal P}_{B}(k,\eta) = \sum_{\lambda=1}^2 \frac{k^4}{4\pi^2 a^4} |A_{k,\lambda}(\eta)|^2, \\
\label{PE} && {\mathcal P}_{E}(k,\eta) = \sum_{\lambda=1}^2 \frac{k^2}{4\pi^2 a^4} |\dot{A}_{k,\lambda}(\eta)|^2.
\end{eqnarray}
Defining also the electromagnetic energy density spectrum, $\rho_{\rm em}(k,\eta)$,
and the electromagnetic trace spectrum, $T_{\rm em}(k,\eta)$, through
\begin{eqnarray}
\label{Espectrum}
&& \rho_{\rm em}(\eta) = \int_0^{\infty} \! \frac{dk}{k} \, \rho_{\rm em}(k,\eta), \\
\label{Tspectrum}
&& T_{\rm em}(\eta) = \int_0^{\infty} \! \frac{dk}{k} \, T_{\rm em}(k,\eta),
\end{eqnarray}
we recast Eqs.~(\ref{TS1}) and~(\ref{TS2}) as
\begin{eqnarray}
\label{T2new}
\rho_{\rm em}(k,\eta) \!\!& = &\!\! \frac12 \, \tau_1 {\mathcal P}_{E} + \frac12 \, \tau_2 {\mathcal P}_{B}, \\
\label{T3new}
T_{\rm em}(k,\eta) \!\!& = &\!\! \tau_3 {\mathcal P}_{E} + \tau_4 {\mathcal P}_{B},
\end{eqnarray}
where we have defined
\begin{eqnarray}
\label{chi1}
&&
\tau_1 = \varepsilon - 2(\xi_1 - \xi_2) f(\phi) \rho_K, \\
\label{chi2}
&&
\tau_2 = \mu^{-1} + 2(\xi_1 + \xi_2) f(\phi) \rho_K, \\
\label{chi3}
&&
\tau_3 = \varepsilon - f(\phi), \\
\label{chi4}
&&
\tau_4 = f(\phi) - \mu^{-1}.
\end{eqnarray}
Let us now specialize our results to the case of de~Sitter spacetime and for large-scale, super-horizon modes.
Inserting the asymptotic expansion for $-k\eta \rightarrow 0$
of the solution~(\ref{p3}) in Eqs.~(\ref{PB}) and~(\ref{PE}), we get
\begin{eqnarray}
\label{P1} && {\mathcal P}_{B}(k,\eta) = \frac{|c_\nu|^2}{4\pi} \,  Z n^4 (-\omega_k\eta)^{5+2\nu} H^4, \\
\label{P2} && {\mathcal P}_{E}(k,\eta) = \frac{4\nu^2}{(-k\eta)^2} \, {\mathcal P}_B(k,\eta),
\end{eqnarray}
respectively. We are principally interested in the case of a scaling-invariant
magnetic spectrum (the general case goes along the same lines as below), so that we take
$\nu = -5/2$ [corresponding to $\gamma = 6$ in Eq.~(\ref{geta})]. 
In this case, we have
\begin{eqnarray}
\label{P3} && {\mathcal P}_{B}(k,\eta) = \frac{9}{2\pi^2} \, Z n^4 H^4, \\
\label{P4} && {\mathcal P}_{E}(k,\eta) = \frac{225}{2\pi^2} \, Z n^4 \, \frac{H^4}{(-k\eta)^2} \, .
\end{eqnarray}
Looking at Eqs.~(\ref{T2new})-(\ref{T3new}) and Eqs.~(\ref{P3})-(\ref{P4}),
and observing that $Z = n/\varepsilon \sim 1/f(\phi)$, we conclude, following the discussion in section~IIIb,
that the electromagnetic backreaction on inflation is not generally negligible. This conclusion could be avoided
if the coefficients $\tau_1$ and $\tau_3$, which enter in the definition of the electric part of
the electromagnetic energy-momentum tensor, 
are vanishing. This happens only if the background tensor $(k_F)^\mu_\nu$ assumes a particular form,
which we are now going to determine.
Assuming that $\tau_3 = 0$, we straightforwardly get
\begin{equation}
\label{ef} \varepsilon = f(\phi), \;\;\; \mu = \frac{n^2}{f(\phi)} \, , \;\;\; Z = \frac{n}{f(\phi)} \, .
\end{equation}
The above equations, when combined with the condition $\tau_1 = 0$ and Eq.~(\ref{g1}), give
\begin{equation}
\label{rhoK} \rho_K = \frac{1}{2(\xi_1 - \xi_2)} \, , \;\;\; p_K = \frac{1}{2(3\xi_1 + \xi_2)} \, ,
\end{equation}
which determine the form of $(k_F)^\mu_\nu$ in Eq.~(\ref{kmunu}) as a function
of $\xi_i$ and, accordingly,
\begin{equation}
\label{nn} n = \sqrt{\frac{(\xi_1 - \xi_2)(3\xi_1 + \xi_2)}{3(\xi_1 - \xi_2)^2 - 4\xi_2^2}} \, .
\end{equation}
Imposing the reality condition $n^2 > 0$, we find
\begin{equation}
\label{reality}
\xi_1 = 0 \;\;\; \mbox{or} \;\;\; \xi_2 = 0 \;\;\; \mbox{or} \;\;\; \frac{\xi_1}{\xi_2} \in \mathbb{X},
\end{equation}
where
\begin{eqnarray}
\label{realityX}
\mathbb{X} \!\!& = &\!\!
\left(-\infty, -\frac13 \right) \cup \left(1 - \frac{2}{\sqrt{3}} \, , 0 \right) \cup \left(0 , 1 \right) \nonumber \\
\!\!& \cup &\!\! \left(1 + \frac{2}{\sqrt{3}} \, , +\infty \right) \! .
\end{eqnarray}
Taking into account Eqs.~(\ref{ef}) and (\ref{rhoK}), we find
\begin{eqnarray}
\label{P5} {\mathcal P}_{B}(k,\eta) \!\!& = &\!\! \frac{9n^5}{2\pi^2 \! f(\phi)} \, H^4, \\
\label{T3} \rho_{\rm em}(k,\eta) \!\!& = &\!\! \frac{9n^5 \Upsilon_1}{2\pi^2} \, H^4, \\
\label{T4} T_{\rm em}(k,\eta) \!\!& = &\!\! \frac{9n^5 \Upsilon_2}{2\pi^2} \, H^4,
\end{eqnarray}
where we have defined
\begin{eqnarray}
\label{T5}
\Upsilon_1 \!\!& = &\!\! \frac{\xi_1(3\xi_1 - \xi_2)}{(\xi_1 - \xi_2)(3\xi_1 + \xi_2)} \, , \\
\label{T6}
\Upsilon_2 \!\!& = &\!\! \frac{4\xi_1 \xi_2}{(\xi_1 - \xi_2)(3\xi_1 + \xi_2)} \, .
\end{eqnarray}
We observe that all the three of the above spectra are scaling-invariant,
and that $\rho_{\rm em}(k,\eta)$ and $T_{\rm em}(k,\eta)$ are time-independent, while the
time dependence of ${\mathcal P}_{B}(k,\eta)$ is all encoded in $f(\phi)$.

Finally, imposing that $\langle (T_{\rm em})^\mu_\nu \rangle \ll (T_{\rm inf})^\mu_\nu$, we get
\begin{eqnarray}
\label{T3bis}
\frac{|\rho_{\rm em}|}{\rho_{\rm inf}} \!\!& = &\!\! 32 \, n^5 |\Upsilon_1| \left(\frac{M}{\mPl}\right)^{\!\!4} \ll 1, \\
\label{T4bis}
\frac{|T_{\rm em}|}{\rho_{\rm inf}} \!\!& = &\!\! 32 \, n^5 |\Upsilon_2| \left(\frac{M}{\mPl}\right)^{\!\!4} \ll 1,
\end{eqnarray}
which are the wanted conditions that must be satisfied in order to have a negligible electromagnetic
backreaction on inflationary dynamics.

\section{VII. Actual magnetic field}

We have seen that inflation is able to produce super-horizon magnetic field fluctuations whose intensity
is given by the magnetic power spectrum~(\ref{P5}). For the following discussion, it is useful to define
the magnetic field strength on the scale $\lambda = 1/k$ as
\begin{equation}
\label{Bl} B(\lambda,\eta) = \sqrt{{\mathcal P}_{B}(1/\lambda,\eta)}.
\end{equation}
At the end of inflation, we have then the scale-invariant magnetic field
\begin{equation}
\label{Bend} B_{\rm end} = \frac{3}{\sqrt{2}\pi} \left(\frac{n^5}{f_{\rm end}}\right)^{\!1/2} H^2,
\end{equation}
where $B_{\rm end} = B(\lambda,\eta_{\rm end})$ and $f_{\rm end} = f(\phi(\eta_{\rm end}))$.

Such a field will evolve from the end of inflation until today. In this section, we will
find the actual value of the inflation-produced magnetic field as a function of the
free parameters of the model, namely the constants $\xi_1$ and $\xi_2$
[the other two free parameters, $\rho_K$ and $p_K$, are assumed to be fixed by Eq.~(\ref{rhoK})]
and, consequently, find the regions in the parameter space $(\xi_1,\xi_2)$ where it satisfies both the
constraint in Eq.~(\ref{Bseed}) and those in Eqs.~(\ref{T3bis})-(\ref{T4bis}).

\subsection{VIIa. Evolution after reheating}

In order to find the present intensity of the magnetic field, we must evolve
it from the end of inflation until the present time $\eta_0$.
As in section IIIb, and to simplify the analysis, we consider the case
of instantaneous reheating.
After the end of reheating (which corresponds in this case to the end of inflation and the
beginning of radiation era), the dynamics of the inflation-produced electromagnetic
field is governed by standard Lagrangian
\begin{equation}
\label{SL} {\mathcal L}_{\rm em} = {\mathcal L}_M + j^\mu A_{\mu},
\end{equation}
since $f(\phi(\eta)) = 1$ after inflation. Here, we have assumed, for the sake of simplicity, that
the background tensor field is vanishingly small for $\eta > \eta_{\rm end}$. This assures that
the experimental constraints on the coefficients of the Lorentz violation $(k_F)_{\mu \nu \alpha \beta}$
are automatically fulfilled~\cite{Data}.

The post-inflationary external electric current $j^\mu$ is vanishing
on superhorizon scales due to causality~\cite{Barrow}, while inside the horizon
it can be written as $j_\mu = (0,-\tilde{\sigma}_c \, \E)$, where
$\tilde{\sigma}_c = a \sigma_c$ is the comoving conductivity and
$\sigma_c$ is the standard conductivity of the plasma. Accordingly,
the equation of the motion for the comoving magnetic field $a^2 \B$,
also known as the magnetic flux $\F$, is
\begin{equation}
\label{flux} \ddot{\F} - \nabla^2 \F = 0
\end{equation}
for modes that live outside the horizon, and~\cite{Turner-Widrow}
\begin{equation}
\label{flux} \ddot{\F} - \nabla^2 \F = -\tilde{\sigma}_c \dot{\F}
\end{equation}
for subhorizon modes. Going into Fourier space,
$\F_\kk(\eta) = \int \! d^3x \, e^{i\kk \x} \F(\x,\eta)$,
and observing that $|k^2 \F_\kk|/|\ddot{\F}_\kk| \sim (-k\eta)^2$,
we find that superhorizon magnetic modes ($-k\eta \ll 1$)
evolve according to $\ddot{\F}_\kk = 0$, so that they scale adiabatically, $\B \propto a^{-2}$.
Modes inside the horizon ($-k\eta \gg 1$), instead,
evolve according to the so-called (comoving) autoinduction equation (see, e.g.,~\cite{Campanelli3}),
$\dot{\F}_\kk = -(k^2/\tilde{\sigma}_c) \F_\kk$.
The solution of the above equation is
\begin{equation}
\label{solflux} \F_\kk(\eta) = \F_\kk(\eta_{\rm RH}) \, e^{-k^2 \ell_d^2/(2\pi)^2} ,
\end{equation}
where
\begin{equation}
\label{elldiss} \ell_d(\eta) = 2\pi \sqrt{\int_{\eta_{\rm RH}}^{\eta} \frac{d\eta'}{\tilde{\sigma}_c(\eta')}}
\end{equation}
is the comoving dissipation length and RH indicates the time of reheating.
Accordingly, modes with wavenumber $k \ll 2\pi/\ell_d$ evolve adiabatically,
while modes with $k \gg 2\pi/\ell_d$ are dissipated.
\footnote{Here, we are neglecting possible effects of magnetohydrodynamic turbulence
that could take place in correspondence of the electroweak and/or quark-hadron (QCD) phase transitions,
and that could affect the evolution of the inflation-generated magnetic
field~\cite{Campanelli3,MHD1,MHD2,MHD3,MHD4,MHD5,MHD6,MHD7,MHD8,MHD11,MHD12,MHD13,MHD14,MHD15,MHD16}.
However, it has been recently shown~\cite{Campanelli4} that a scaling invariant magnetic field
stays almost unchanged on scales of cosmological interest, although on smaller scales its spectrum
is progressively suppressed.}

Putting all together, we conclude that, during the expansion of the Universe after reheating,
magnetic modes are washed out on scales below the dissipation length and diluted adiabatically
on larger scales. However, the actual dissipation length is very small compared to the scale of
interest for cosmic magnetic fields.

To see this, we firstly remember the conductivity $\sigma_c$ depends, generally,
on the temperature $T$~\cite{Turner-Widrow}.
In the radiation-dominated era, and for temperatures much greater than the electron mass $m_e$,
we have $\sigma_c(T) \sim T/e^2$~\cite{Turner-Widrow}, where $e$ is the absolute value of the electric charge.
After the epoch of $e^+ e^-$ annihilation ($T_{\rm anh} \sim m_e$), the conductivity
is given by $\sigma_c(T) \sim (T/e^2)\sqrt{T/m_e}$~\cite{Dimopoulos-Davis},
while in the matter-dominated era ($T < T_{\rm eq} \simeq 3\eV$), and after electrons and ions recombine
($T_{\rm rec} \simeq 0.3\eV$), it drops to the constant value
$\sigma_c(T) \sim 10^{-13} m_e/e^2 \simeq 8 \times 10^{8} s^{-1}$~\cite{Turner-Widrow}.
Taking into account that $\eta \propto a$ and $\eta \propto a^{1/2}$ in the radiation-dominated
and matter-dominated eras, respectively, and that
$a \propto g_{*S}^{-1/3} T^{-1}$ after reheating, where $g_{*S}(T)$ is the effective
number of entropy degrees of freedom at the temperature $T$~\cite{Kolb-Turner},
we conveniently split the integral in Eq.~(\ref{elldiss0}), evaluated at the present time, in four integrals,
$\int_{\eta_{\rm RH}}^{\eta_0} d\eta/\tilde{\sigma}_c = I_1 + I_2 + I_3 + I_4$.
Here,
$I_1 = \int_{\eta_{\rm RH}}^{\eta_{\rm anh}} d\eta/\tilde{\sigma}_c$,
$I_2 = \int_{\eta_{\rm anh}}^{\eta_{\rm eq}} d\eta/\tilde{\sigma}_c$,
$I_3 = \int_{\eta_{\rm eq}}^{\eta_{\rm rec}} d\eta/\tilde{\sigma}_c$,
and
$I_4 = \int_{\eta_{\rm rec}}^{\eta_0} d\eta/\tilde{\sigma}_c$.
Since $I_1/I_2 \sim (T_{\rm eq}/m_e)^{3/2} \sim 10^{-8}$,
$I_2/I_3 \sim T_{\rm rec}/T_{\rm eq} \sim 10^{-1}$, and
$I_3/I_4 \sim 10^{-13} (m_e/T_{\rm rec})^{3/2} \sim 10^{-4}$,
the integral $I_4$ dominates over the other three in the expression for the
actual dissipation length. Accordingly, we have
\begin{equation}
\label{elldiss0} \ell_d(t_0) \simeq
2\pi \left(\frac{3t_0}{\sigma_c}\right)^{\!\!1/2} \! \left(\frac{t_0}{t_{\rm rec}}\right)^{\!\!1/6}
\simeq 10^{-2} \pc,
\end{equation}
where we used the fact that in the matted-dominated era $a(t) \simeq (t/t_0)^{2/3}$ and then
$\eta(t) \simeq 3t_0 (t/t_0)^{1/3}$, and $t_{\rm rec} \simeq 8 \times 10^{12} s$~\cite{Kolb-Turner}.

\subsection{VIIb. Actual magnetic field strength}

As anticipated, the actual dissipation length is negligibly small compared to the scale of
interest for cosmic magnetic fields, which is of order of $1 \Mpc$.
We conclude that the inflation-produced magnetic field evolve adiabatically, from the time of reheating until today.
Its actual intensity is then
\begin{equation}
\label{Bk0} B_0 = B_{\rm end} \left(\frac{g_{*S,0}}{g_{*S,{\rm RH}}} \right)^{\!\!2/3} \!
\left(\frac{T_0}{T_{\rm RH}}\right)^{\!2} \cos \theta_{W},
\end{equation}
where $B_0 = B(\lambda,\eta_0)$, $g_{*S,0} = g_{*S}(T_0) = 43/11$~\cite{Olive},
$g_{*S,\rm{RH}} = g_{*S}(T_{\rm RH})$~\cite{Olive}, $T_0 \simeq 2.37 \times 10^{-4} \eV$~\cite{Kolb-Turner}
is the actual temperature, and $T_{\rm RH}$ is the reheat temperature.
Above the electroweak phase transition (when we
assume inflation is taking place) the $U(1)$ gauge field which is quantum mechanically
excited is indeed the hypercharge field, not the electromagnetic one~\cite{Ratra}. Below the electroweak
phase transition, however, the hypercharge field is projected onto the electromagnetic field,
and this gives the cosine of the Weinberg angle $\theta_{W}$.

The reheat temperature can be related to the energy scale of inflation by observing that the
energy density of radiation at the beginning of radiation era,
$\rho_{\rm rad} = (\pi^2/30) g_{*,\rm{RH}} \, T_{\rm RH}^4$,
where $g_{*,\rm{RH}}$ is the effective number of degrees of freedom at the time of reheating
and can be taken equal to $g_{*S,\rm{RH}}$~\cite{Kolb-Turner}, must be equal to the energy
density at the end of inflation. We get $T_{\rm RH} = [30/(\pi^2 g_{*,\rm{RH}})]^{1/4} M$.
Taking $g_{*S,\rm{RH}} = 427/4$~\cite{Kolb-Turner}, referring to the massless degrees of
freedom of the standard model of particle physics above the electroweak scale,
the actual, scale-invariant magnetic field is
\begin{equation}
\label{B0} B_0 \simeq 2 \times 10^{-12} \left(\frac{n^5}{f_{\rm end}}\right)^{\!\!1/2}
\! \left(\frac{M}{10^{16}\GeV}\right)^{\!\!2} \G.
\end{equation}
Let us now take $f_{\rm end} \sim 1$ and $M \sim 10^{16} \GeV$. Accordingly, we have
\begin{equation}
\label{B0bis} B_0 \sim n^{5/2} 10^{-12} \G.
\end{equation}
The condition that the electromagnetic backreaction on inflation is negligible,
expressed by Eqs.~(\ref{T3bis}) and (\ref{T4bis}), becomes
\begin{eqnarray}
\label{T3tris}
&& n^5 |\Upsilon_1| \ll 10^{11} , \\
\label{T4tris}
&& n^5 |\Upsilon_2| \ll 10^{11}.
\end{eqnarray}
Let us now analyze, separately, the three cases in Eq.~(\ref{reality}), namely
$\xi_1 = 0$, $\xi_2 = 0$, and $\xi_1/\xi_2 \in \mathbb{X}$.

($i$) $\xi_1 = 0$. -- This corresponds, looking at Lagrangian density~(\ref{Lagrangian2bis}),
to the case where the electromagnetic field is coupled only to the traceless part of the background
tensor $(k_F)^\mu_\nu$. From Eqs.~(\ref{rhoK}), (\ref{nn}), (\ref{T3}), and~(\ref{T4}), we get
$\rho_K = -1/2\xi_2$, $p_K = -\rho_K$, 
$n=1$, $\Upsilon_1 = 0$, and $\Upsilon_2 = 0$, respectively.
This gives, in turns, $\langle (T_{\rm em})^\mu_\nu \rangle = 0$,
so that the electromagnetic backreaction on inflation is absent.
The actual, scaling-invariant magnetic field is
of order of $B_0 \sim 10^{-12} \G$ and it can directly account for the presently observed cosmic magnetic fields.

($ii$) $\xi_2 = 0$. -- This case corresponds to an
electromagnetic field coupled only to the scalar part (trace) of $(k_F)^\mu_\nu$.
We have $\rho_K = 1/2\xi_1$, $p_K = \rho_K/3$, 
$n = 1$, $\Upsilon_1 = 1$, and $\Upsilon_2 = 0$ (which gives $T_{\rm em} = 0$). 
The electromagnetic backreaction on inflation is, then, completely negligible
and, also in this case, the inflation-produced magnetic field can directly explain
cosmic magnetization.

($iii$) $\xi_1/\xi_2 \in \mathbb{X}$. -- In Fig.~1, we plot the function $n^{5/2}$, entering in the expression
of $B_0$ in Eq.~(\ref{B0bis}), at the varying of $\xi_1/\xi_2$.
Remembering the discussion in section~II [see, in particular, Eq.~(\ref{Bseed})], we find that,
in order to explain cosmic magnetization, the quantity $n^{5/2}$ should be in the range $[\, 0.1, few]$.
This is realized for all values of $\xi_1/\xi_2 \in \mathbb{X}$, with the exclusion of those
values very close to the boundary
$\partial \mathbb{X} = \{-1/3\} \cup \{1-2/\sqrt{3}\} \cup \{1\} \cup \{1+2/\sqrt{3}\}$.
In fact, $n$ diverges for $\xi_1/\xi_2 \rightarrow -1/3$
and $\xi_1/\xi_2 \rightarrow 1+2/\sqrt{3}$, while it goes to zero for
$\xi_1/\xi_2 \rightarrow 1-2/\sqrt{3}$ and $\xi_1/\xi_2 \rightarrow 1$.
When $\xi_1/\xi_2$ is not so close to the above boundary values,
the conditions~(\ref{T3tris}) and (\ref{T4tris}), which assure that the
electromagnetic field does not back-react on the inflationary dynamics,
are fulfilled. This is clear from Fig.~1, where we show $n^5 \Upsilon_1$
and $n^5 \Upsilon_2$ at the varying of $\xi_1/\xi_2$.
We conclude that, apart from some particular values of $\xi_1/\xi_2$, the inflation-produced magnetic
field can be, also in this case, at the origin of cosmic magnetic fields.


\begin{figure*}[t!]
\begin{center}
\hspace{-0.5cm}
\includegraphics[scale=0.37,bb=0 0 600 600]{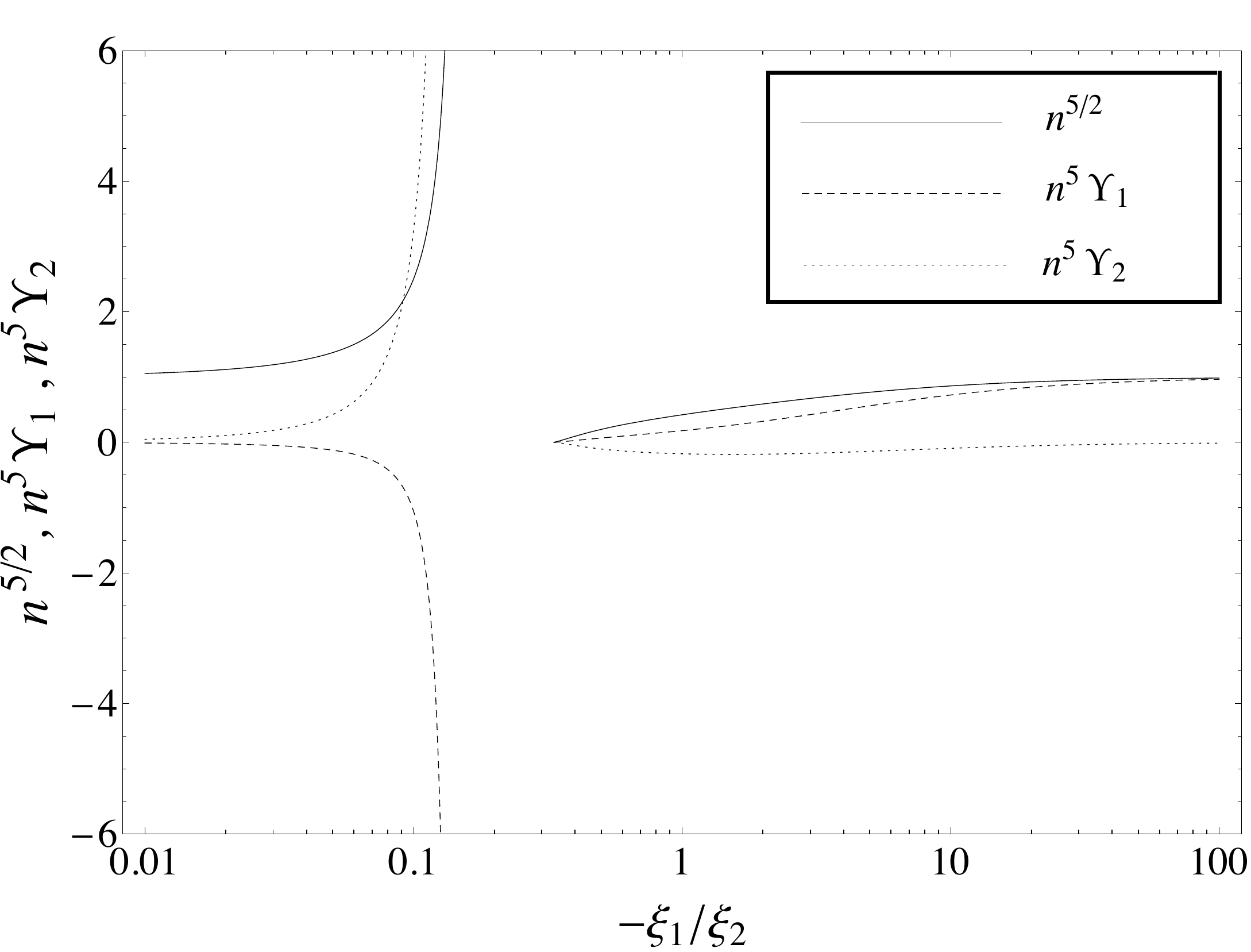}
\hspace{0.8cm}
\includegraphics[scale=0.37,bb=0 0 600 600]{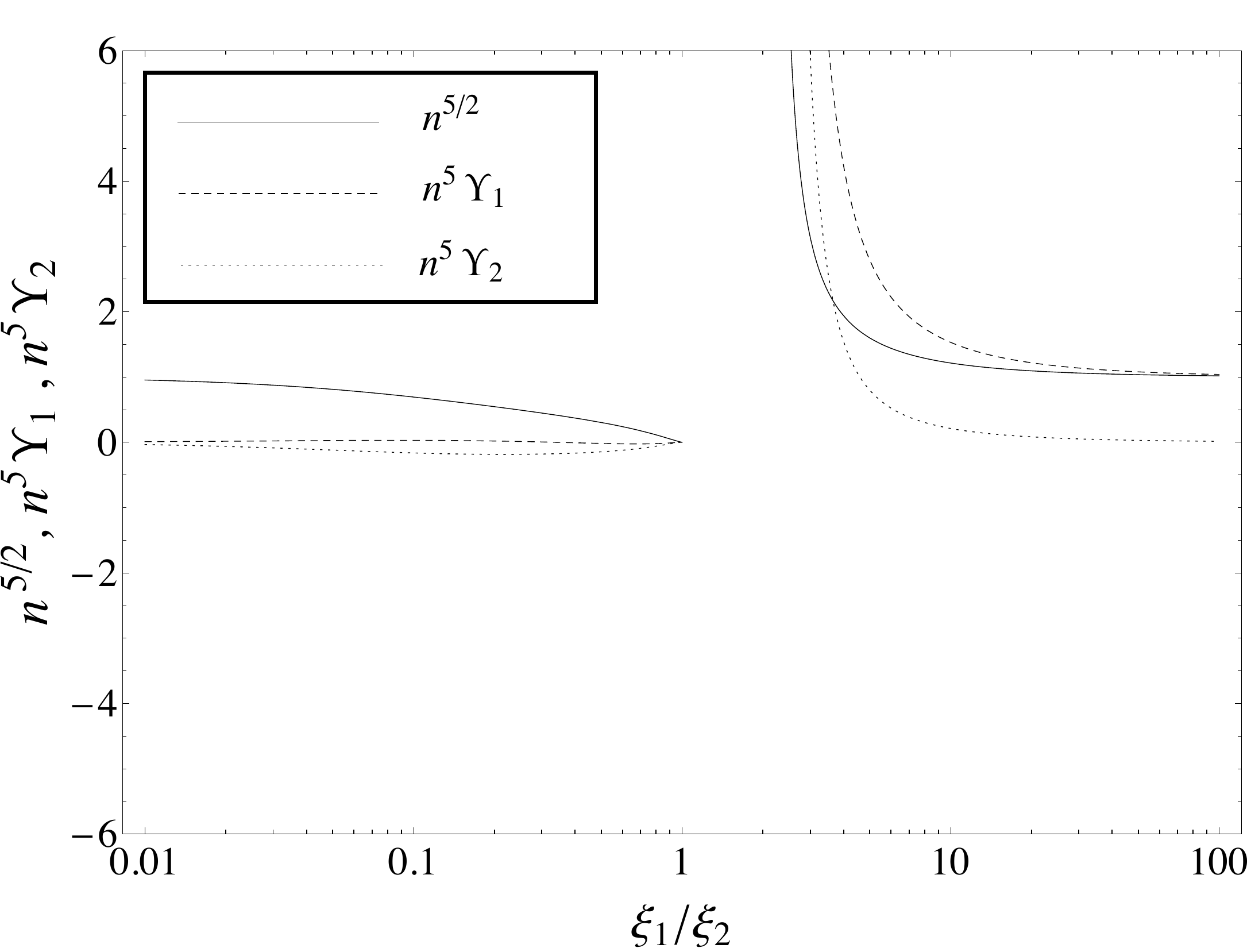}
\caption{The quantities $n^{5/2}$ (continuous lines), $n^5 \Upsilon_1$ (dashed lines),
and $n^5 \Upsilon_2$ (dotted lines) [which appear in
Eqs.~(\ref{B0bis}), (\ref{T3tris}), and (\ref{T4tris}), respectively],
as a function of $\xi_1/\xi_2$.}
\end{center}
\end{figure*}


\section{VIII. Additional conditions on $(T_{\rm em})^\mu_\nu$}


\begin{figure*}[t!]
\begin{center}
\hspace{-0.5cm}
\includegraphics[scale=0.37,bb=0 0 600 600]{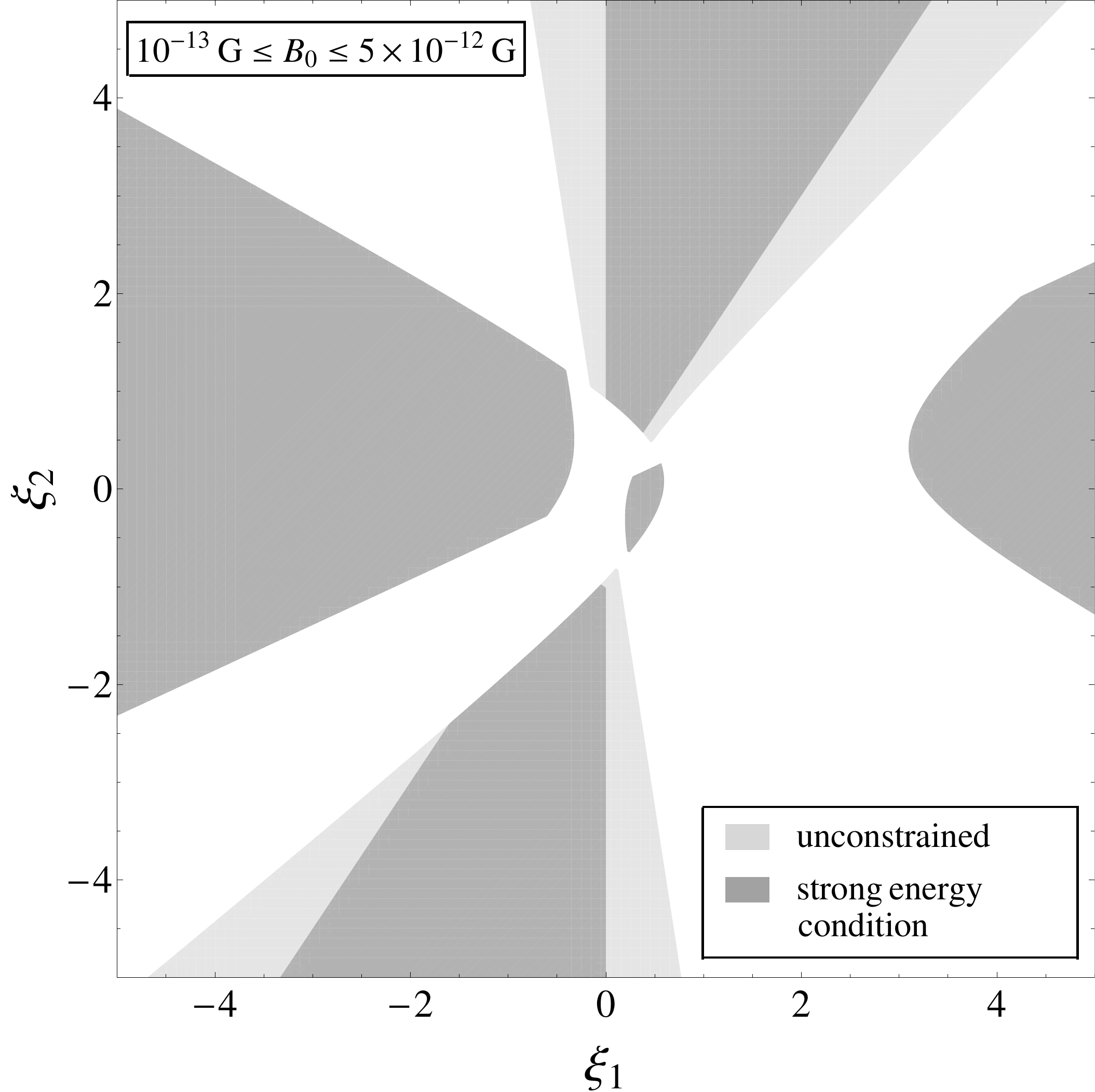}
\hspace{0.345cm}
\includegraphics[scale=0.37,bb=0 0 600 600]{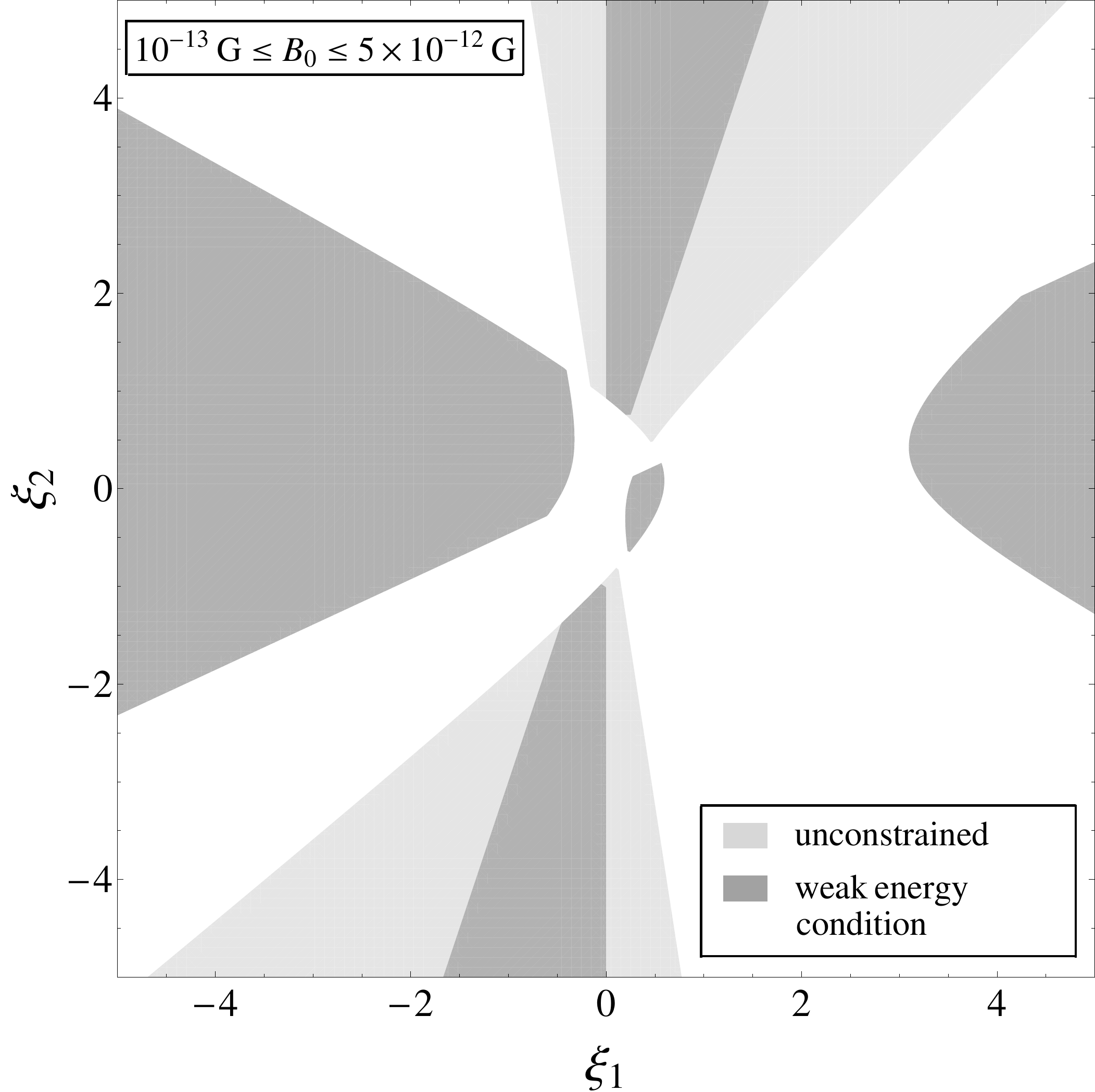}

\vspace{0.6cm}

\hspace{-0.5cm}
\includegraphics[scale=0.37,bb=0 0 600 600]{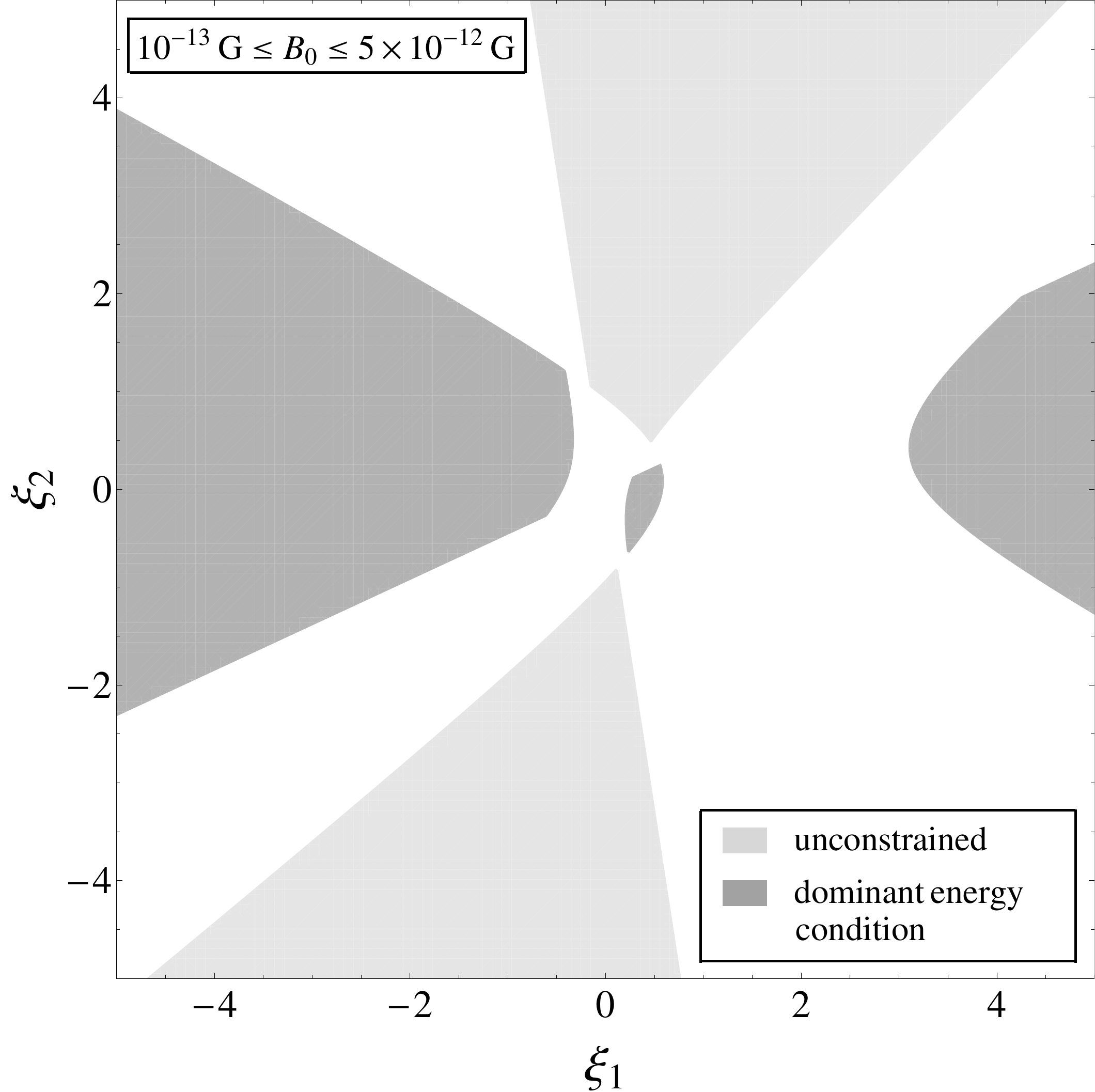}
\hspace{0.345cm}
\includegraphics[scale=0.37,bb=0 0 600 600]{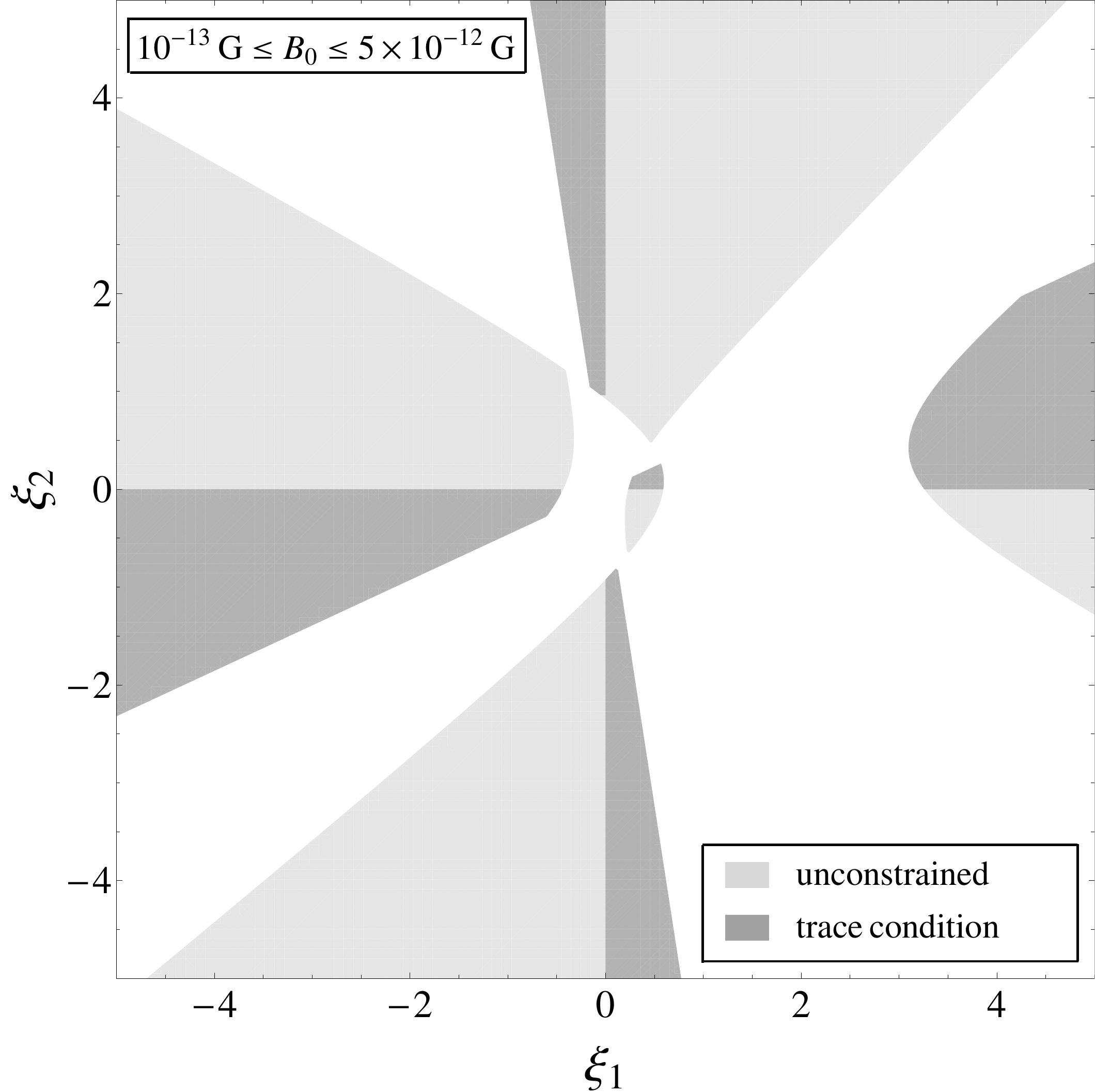}
\caption{Regions in the parameter space $(\xi_1,\xi_2)$, with $\xi_1/\xi_2 \in \mathbb{X}$ [see Eq.~(\ref{realityX})],
where $10^{-13} \G \leq B_0 \leq 5 \times 10^{-12} \G$ and electromagnetic backreaction on
inflation is completely negligible (light gray areas).
In the shrunk dark gray regions, a specific supplementary condition on the
electromagnetic energy-momentum has been imposed (from up to down and from left to right:
strong energy condition, weak energy condition, dominant energy condition, and trace condition).}
\end{center}
\end{figure*}


Let us now impose some physically ``reasonable'' conditions on the inflation-produced electromagnetic
energy-momentum tensor. Some of these conditions, as the positivity of the energy,
are often assumed to be ``necessary'' in the literature.
It is worth noticing, however, that there exist examples of physically ``reasonable'' matter
that violate some or of all of them.
For example, all the conditions that we are going to discuss are violated in particular setups
of the Casimir effect~\cite{Birrell-Davies}, and even the inflaton violates the strong energy condition
(discussed below) when it drives de Sitter inflation.

{\it Weak energy condition.} -- Looking at the left panel of Fig.~1, we see that the quantity $n^5 \Upsilon_1$ is
negative for $\xi_1/\xi_2 < -1/3$. This corresponds to have a negative electromagnetic energy density
on large super-horizon scales during inflation [see Eq.~(\ref{T3})]. On these scales,
we expect that the electromagnetic field behaves classically, so that one could wonder if
having classical negative energies is reasonable physically.
Let us then impose the condition of positivity of the energy. In a general-covariant formulation,
this condition is known as ``weak energy condition'' and it is, indeed, a condition on the energy-momentum tensor.
In our specific case, the electromagnetic energy-momentum tensor can be written as
\begin{equation}
\label{emfluid}
\langle (T_{\rm em})^\mu_\nu \rangle = \mbox{diag} (\rho_{\rm em},-p_{\rm em},-p_{\rm em},-p_{\rm em}).
\end{equation}
This is the energy-momentum tensor of a (isotropic) perfect fluid of type I
(according to the Hawking-Ellis classification~\cite{Hawking-Ellis}) with energy density $\rho_{\rm em}$ and
pressure density $p_{\rm em} = -\langle (T_{\rm em})^i_i \rangle$ (no sum on $i$).
For perfect fluids, the weak energy condition states that~\cite{Hawking-Ellis}
\begin{equation}
\label{ultimo1}
\rho_{\rm em} \geq 0, \;\;\; \rho_{\rm em} + p_{\rm em} \geq 0.
\end{equation}
These supplementary conditions, if applied to Eq.~(\ref{emfluid}), would narrow the domain~(\ref{realityX}) to
$\mathbb{X} =
\left(-\infty, -1/3 \right) \cup \left( 0, 1/3 \right) \cup \left(1 + 2/\sqrt{3} , +\infty \right)$.

{\it Trace condition.} -- It is well known that the trace of the energy-momentum tensor for a system of point-like,
electromagnetic interacting particles is non-negative~\cite{Landau}. This condition is sometimes assumed to be valid
also for other interacting systems in Nature~\cite{Landau}. If we require that
\begin{equation}
\label{ultimo2}
T_{\rm em} \geq 0,
\end{equation}
the domain~(\ref{realityX}) would reduce to
$\mathbb{X} = \left(1 - 2/\sqrt{3}, 0 \right) \cup \left(1 + 2/\sqrt{3}, +\infty \right)$.

There are other restrictions on the energy-momentum tensor
conjectured to hold for all physically reasonable matter. Those are
the 
strong and dominant energy conditions.
(The physical significance of these conditions is, respectively, and roughly speaking,
that matter must gravitate toward matter, and that
energy must either be non-negative and not flow faster than light~\cite{Hawking-Ellis}.)


{\it Strong energy condition.} -- This condition requires that~\cite{Hawking-Ellis}
\begin{equation}
\label{ultimo3}
\rho_{\rm em} \geq 0, \;\;\; \rho_{\rm em} + 3p_{\rm em} \geq 0,
\end{equation}
and would reduce the domain~(\ref{realityX}) to
$\mathbb{X} = \left(-\infty, -1/3 \right) \cup \left( 0, 2/3 \right) \cup \left(1 + 2/\sqrt{3}, +\infty \right)$.

{\it Dominant energy condition.} -- This condition imposes~\cite{Hawking-Ellis}
\begin{equation}
\label{ultimo4}
\rho_{\rm em} \geq |p_{\rm em}|
\end{equation}
and, if applied, it would shrink the domain~(\ref{realityX}) to
$\mathbb{X} = \left(-\infty, -1/3 \right) \cup \left(1 + 2/\sqrt{3}, +\infty \right)$.

If we impose simultaneously all the above conditions,
we would reduce the domain~(\ref{realityX}) to
$\mathbb{X} = \left(1 + 2/\sqrt{3}, +\infty \right)$.
This means that the only ``surviving'' part of Fig.~1 would be the right branch in its right panel.

The light gray areas in Fig.~2 show the regions in the parameter space $(\xi_1,\xi_2)$
where $10^{-13} \G \leq B_0 \leq 5 \times 10^{-12} \G$ 
and electromagnetic backreaction on inflation is completely negligible.
The dark gray areas represent, instead, the shrunk regions where a specific supplementary condition on the
electromagnetic energy-momentum has been imposed (from up to down and from left to right:
strong energy condition, weak energy condition, dominant energy condition, and trace condition).

\section{IX. Curvature perturbations}

Recently enough, it has been pointed out in the literature that the production of
electromagnetic fields during inflation may significantly affect the primordial spectrum
of both scalar and tensor curvature perturbations (see, e.g.,~\cite{Barnaby,Suyama,Barnaby2,Fujita2}).
In order to have a self-consistent model of inflationary magnetogenesis, then,
we have to check that the curvature perturbations introduced by the inflation-produced
electromagnetic field are compatible with CMB results.

\subsection{IXa. Scalar curvature perturbation}

In order to find how a primordial magnetic field can generate curvature perturbations,
let us consider the curvature perturbation $\zeta(t,\x)$ on the uniform energy density
hypersurface~\cite{Lyth} on which $\delta \rho(t,\x) = 0$, where $\delta \rho$ is the energy density perturbation,
and $t$ is the cosmic time.
The curvature perturbation, as a function of the scale factor $a(t,\x)$, is~\cite{Lyth}
\begin{equation}
\label{zetadef} \zeta(t,\x) = \ln a(t,\x) - \ln a(t),
\end{equation}
where $a(t)$ is the global scale factor, namely the one introduced in the unperturbed metric~(\ref{line}).
On super-Hubble scales, the curvature perturbation evolves according to~\cite{Fujita2}
\begin{equation}
\label{zetaeq} \zeta'(t,\x) = - \frac{\delta p_{\rm rel}(t,\x)}{\rho(t) + p(t)} \, H(t),
\end{equation}
where a prime denotes differentiation with respect to the cosmic time.
Here, $\rho(t)$ and $p(t)$ are the total energy and pressure densities, $H(t)$ the Hubble parameter, and
$\delta p_{\rm rel}(t,\x)$ is the so-called nonadiabatic pressure density perturbation defined by
$\delta p_{\rm rel}(t,\x) = \delta p(t,\x) - \delta \rho(t,\x) p'/\rho'$,
with $\delta p$ being the pressure density perturbation.
Assuming that the electromagnetic field is just a small perturbation with respect to the
background (which is dominated by the inflaton field), we can write
$\rho = \rho_{\rm inf}$ and $p = p_{\rm inf}$. The evolution equation for the curvature perturbation
introduced by the electromagnetic field, $\zeta^{\rm em}(t,\x)$, is then
\begin{equation}
\label{zetaeqem} (\zeta^{\rm em})'(t,\x) = - \frac{\delta p_{\rm rel,em}(t,\x)}{\rho_{\rm inf}(t) + p_{\rm inf}(t)} \, H(t),
\end{equation}
where $\delta p_{\rm rel,em}(t,\x) = \delta p_{\rm em}(t,\x) - \delta \rho_{\rm em}(t,\x) p_{\rm inf}'/\rho_{\rm inf}'$
is the nonadiabatic pressure perturbation due to the relative entropy perturbation between the electromagnetic and
the inflaton fields. Here, $\delta \rho_{\rm em}$ and $\delta p_{\rm em}$ are the electromagnetic energy
and pressure density perturbations, respectively, and they are the same quantities defined
in Eq.~(\ref{emfluid}), to wit, $\delta \rho_{\rm em} = \rho_{\rm em}$ and $\delta p_{\rm em} = p_{\rm em}$.

Assuming a quasi-de Sitter inflation characterized by the slow-roll parameter $\epsilon \ll 1$,
$p_{\rm inf} = (1-2\epsilon/3) \rho_{\rm inf}$, and introducing the
electromagnetic equation-of-state $w_{\rm em} = \rho_{\rm em}/p_{\rm em} = \delta \rho_{\rm em}/\delta p_{\rm em}$,
the solution of Eq.~(\ref{zetaeqem}) reads
\begin{equation}
\label{zetaemsol}
\zeta^{\rm em}(t,\x) = -3(1+w_{\rm em}) \, \frac{H}{2\epsilon \rho_{\rm inf}} \int_{t_i}^{t} \! dt' \delta \rho_{\rm em}(t',\x),
\end{equation}
where $t_i$ is the time when electromagnetic fluctuations begin to develop, $\zeta^{\rm em}(t_i,\x) = 0$, and
\begin{equation}
\label{wemx}
w_{\rm em} = \frac{3\xi_1 - 5\xi_2}{3(3\xi_1 - \xi_2)} \, .
\end{equation}
In obtaining Eq.~(\ref{zetaemsol}), we assumed, as in Ref.~\cite{Fujita2}, that
$H$, $\epsilon$, and $\rho_{\rm inf}$ are constant during inflation, and that
$p_{\rm inf}' \simeq \rho_{\rm inf}'$.
Equation~(\ref{wemx}), instead, comes from Eqs.~(\ref{T3}) and (\ref{T4}).
In the case $w_{\rm em} = 1/3$, we recover
the result of Ref.~\cite{Fujita2}.
The curvature perturbation in Eq.~(\ref{zetaemsol}) is the key quantity from which
observable quantities can be constructed and then compared to CMB results.

In Ref.~\cite{Fujita2}, the standard kinetically coupled scenario for magnetogenesis
was studied. In this case, the electromagnetic energy density is dominated by the electric part,
so that, working in Fourier space, the electromagnetic energy spectrum can be approximated by
$\delta \rho_{\rm em}(k,\eta) = \frac12 f(\phi) \mathcal{P}_E(k,\eta)$.
The expression of the electric power spectrum (in the standard kinetically coupled scenario)
can be obtained by using Eqs.~(\ref{P1})-(\ref{P2}), and taking $Z=1/f(\phi)$ and $n = 1$. It is
$\mathcal{P}_E(k,\eta) = (\nu^2 |c_\nu|^2/\pi) (-k\eta)^{3+2\nu} H^4/f(\phi)$. The scaling-invariant case
corresponds to taking $\nu= -3/2$, and it gives $\mathcal{P}_E(k,\eta) = (9/2\pi^2) H^4/f(\phi)$.
Accordingly, we have
\begin{equation}
\label{deltarhoF}
\delta \rho_{\rm em,*}(k,\eta) = \frac{9}{4\pi^2} \, H^4,
\end{equation}
where, here and in the following, a star indicates that the corresponding result is obtained in the
standard kinetically coupled scenario for the case of a scaling-invariant electric power spectrum.

In our case, instead, the electric part does not make any contribution to the electromagnetic energy
which is then dominated by the magnetic part (see discussion in section~VI). The expression for
$\delta \rho_{\rm em}(k,\eta)$ is given in Eqs.~(\ref{T3}), which we rewrite here for the sake of convenience,
\begin{equation}
\label{deltarhonoi}
\delta \rho_{\rm em}(k,\eta) = \frac{9n^5 \Upsilon_1}{2\pi^2} \, H^4.
\end{equation}
Comparing Eqs.~(\ref{deltarhoF}) and (\ref{deltarhonoi}), and taking into account
Eq.~(\ref{zetaemsol}), we see that the curvature perturbation in our case
can be obtained by multiplying the result of Ref.~\cite{Fujita2}
(for the scaling-invariant case) by a constant factor $\vartheta$,
\begin{equation}
\label{zetazeta} \zeta^{\rm em} = \vartheta(\xi_1,\xi_2) \zeta^{\rm em}_*,
\end{equation}
where
\begin{equation}
\label{vartheta}
\vartheta(\xi_1,\xi_2) = \frac{3(1+w_{\rm em})}{2} \, n^5 \Upsilon_1 = n^5 (2 \Upsilon_1 - \Upsilon_2),
\end{equation}
and where, in the last equality of the above equation, we used Eqs.~(\ref{T5}), (\ref{T6}), and (\ref{wemx}).

\subsection{IXb. Scalar modes: spectrum, bispectrum, and trispectrum}

The observable quantities that can be constructed starting from the curvature perturbation $\zeta(t,\x)$,
are the corresponding $n$-points correlation functions in Fourier space. In particular,
the actual sensitivity of CMB experiments allows us to put constraints on the
2-points correlator, the {\it power spectrum} of curvature perturbations,
on the 3-point correlator, the {\it bispectrum},
and on the 4-points correlator, the {\it trispectrum}, defined via
\begin{eqnarray}
\label{2def}
\langle \zeta_{\kk_1} \zeta_{\kk_2} \rangle \!\!& = &\!\!
(2\pi)^3 \delta(\kk_1 + \kk_2) \frac{2\pi^2}{k_1^3} \, \mathcal{P}_\zeta, \\
\label{3def}
\langle \zeta_{\kk_1} \zeta_{\kk_2} \zeta_{\kk_3} \rangle \!\!& = &\!\!
(2\pi)^3 \delta(\kk_1 + \kk_2 + \kk_3) (2\pi^2\mathcal{P}_\zeta)^2 \nonumber \\
\!\!& \times &\!\! \frac65 \, f_{\rm NL}^{\rm local}
\frac{\sum_{i=1}^3 k_i^3}{\prod_{i=1}^3 k_i^3} \, , \\
\label{4def}
\langle \zeta_{\kk_1} \zeta_{\kk_2} \zeta_{\kk_3} \zeta_{\kk_4} \rangle \!\!& = &\!\!
(2\pi)^3 \delta(\kk_1 + \kk_2 + \kk_3 + \kk_4) (2\pi^2\mathcal{P}_\zeta)^3 \nonumber \\
\!\!& \times &\!\! \tau_{\rm NL} \! \left[ \frac{1}{(k_1 k_2 k_{13})^3} + 11 \: \mbox{permutations} \right] \!, \nonumber \\
\end{eqnarray}
respectively, where a scaling-invariant power spectrum $\mathcal{P}_\zeta(k)$ is assumed, and all quantities
are evaluated at the end of inflation $\eta = \eta_{\rm end}$.
Here, $\zeta_{\kk}$ is the Fourier-transformed curvature perturbation,
$\langle ... \rangle$ indicates an ensemble average,
$k_i = |\kk_i|$, and $\kk_{ij} = |\kk_i + \kk_j|$.
The observable quantities, besides $\mathcal{P}_\zeta(k)$,
are the local-type non-linearity (dimensionless) parameters $f_{\rm NL}^{\rm local}$ and $\tau_{\rm NL}$,
which parameterize the non-Gaussian features of the primordial spectrum of curvature perturbations.

Let us observe that the 3-points correlator can be generally written as
$\langle \zeta_{\kk_1} \zeta_{\kk_2} \zeta_{\kk_3} \rangle =
(2\pi)^3 \delta(\kk_1 + \kk_2 + \kk_3) f_{\rm NL} \, \mathcal{P}_\zeta(k_1,k_2,k_3)$,
where the bispectrum $\mathcal{P}_\zeta(k_1,k_2,k_3)$ measures the correlation among
three perturbation modes~\cite{Planck1}. The bispectrum can assume different forms
which depend on the type of triangle formed by the three wavenumbers
$k_1$, $k_2$ and $k_3$. Local-type non-linearities are characterized
by a bispectrum that is maximal for ``squeezed''
triangles with $k_1 \ll k_2 \simeq k_3$ (or permutations).
Other types of configurations are possible, such as the equilateral
and the orthogonal. They are, however, inessential for our discussion since
a scaling-invariant electromagnetic field produces (under appropriate approximations~\cite{Fujita2})
only local-type shapes for the bispectrum and trispectrum.

Using the results of Ref.~\cite{Fujita2}, appropriately re-scaled by using Eq.~(\ref{zetazeta}),
we find for the electromagnetic part of the power spectrum of curvature perturbations, $\mathcal{P}_\zeta^{\rm em}$,
and for the electromagnetic part of the local-type non-linearity parameters, $f_{\rm NL}^{\rm em}$ and $\tau_{\rm NL}^{\rm em}$,
\begin{eqnarray}
\label{zeta} && \mathcal{P}_\zeta^{\rm em} =
192 \, \vartheta^2 N_{\rm cmb}^2 \Delta N (\mathcal{P}_\zeta^{\rm inf})^2,
\\
\label{zeta2} && f_{\rm NL}^{\rm em} =
\frac{20}{3} \, \vartheta N_{\rm cmb} \, \frac{\mathcal{P}_\zeta^{\rm em}}{\mathcal{P}_\zeta^{\rm inf}} \, ,
\\
\label{zeta3} && \tau_{\rm NL}^{\rm em} =
72 \, \vartheta^2 N_{\rm cmb}^2 \, \frac{\mathcal{P}_\zeta^{\rm em}}{\mathcal{P}_\zeta^{\rm inf}} \, ,
\end{eqnarray}
respectively. Here, it has been assumed that the dominant component of the power spectrum of
curvature perturbations is generated by the inflaton,
\begin{equation}
\label{PemPinf}
\mathcal{P}_\zeta^{\rm em} \ll \mathcal{P}_\zeta^{\rm inf},
\end{equation}
where the power spectrum of curvature perturbations in slow-roll inflation is~\cite{Fujita2}
\begin{equation}
\label{Pinf} \mathcal{P}_\zeta^{\rm inf} = \frac{1}{24\pi^2\epsilon} \left(\frac{M}{\mPl}\right)^{\!4} \! .
\end{equation}
Moreover, it has been assumed that $\Delta N > 1$, where
\begin{equation}
\label{Delta} \Delta N = N_{\rm tot} - N_{\rm cmb},
\end{equation}
with $N_{\rm tot} = -\ln|k_{\rm min} \eta_{\rm end}|$ and $N_{\rm cmb} = -\ln|k_{\rm cmb} \eta_{\rm end}|$.
Here, $k_{\rm min}$ is the wavenumbers of the mode that crosses the
horizon when magnetogenesis begins, and $k_{\rm cmb}$, referred to as the CMB scale,
is the wavenumber of the largest scale CMB mode.
We have assumed, in the previous sections, that magnetogenesis begins simultaneously with inflation, so that
$k_{\rm min} = -1/\eta_i$, and in turns $N_{\rm tot}$ is the total number of $e$-fold of inflation.
On the other hand, $N_{\rm cmb}$ is the number of $e$-folds
between the moment at which the CMB mode leaves the horizon and the end of inflation.
Assuming for simplicity an instantaneous reheating, we have~\cite{Kolb-Turner}
\begin{equation}
\label{Ncmb} N_{\rm cmb} \simeq 61 + \ln \frac{\lambda_{\rm cmb}}{4000 \Mpc} + \ln \frac{M}{10^{16} \GeV} \, ,
\end{equation}
where $\lambda_{\rm cmb} = 2\pi/k_{\rm cmb}$. Taking $M = 10^{16} \GeV$ and
the CMB scale as large as the the present Hubble radius, $\lambda_{\rm cmb} = H_0^{-1} \simeq 4000 \Mpc$,
we get $N_{\rm cmb} \simeq 61$. %

We use the experimental results derived from Planck data~\cite{Planck0,Planck2,Feng},
\begin{eqnarray}
\label{zetao}
&& \!\!\!\!\!\!\!\!\!\!\! \mathcal{P}_\zeta^{\rm inf} = 2.2 \times 10^{-9} ~~ (\mbox{best-fit}),
\\
\label{zeta2o}
&& \!\!\!\!\!\!\!\!\!\!\! f_{\rm NL}^{\rm local} = 2.5 \pm 5.7 ~~ (68\% \, \mbox{C.L.}),
\\
\label{zeta3o}
&& \!\!\!\!\!\!\!\!\!\!\! \tau_{\rm NL} = (0.3 \pm 0.9) \times 10^4 ~~ (68\% \, \mbox{C.L.}),
\end{eqnarray}
where $\mathcal{P}_\zeta^{\rm inf}(k)$ is evaluated at the pivot scale $k_0 = 0.05 \Mpc^{-1}$.
Taking into account Eq.~(\ref{zetao}), we can conveniently rewrite Eqs.~(\ref{zeta}), (\ref{zeta2}),
and (\ref{zeta3}) as
\begin{eqnarray}
\label{zetaa}
&& \frac{\mathcal{P}_\zeta^{\rm em}}{\mathcal{P}_\zeta^{\rm inf}} \simeq
1.6 \times 10^{-3} \, \vartheta^2 \left(\frac{N_{\rm cmb}}{61}\right)^{\!2} \Delta N,
\\
\label{zeta2a}
&& f_{\rm NL}^{\rm em} \simeq 0.6 \, \vartheta^3 \left(\frac{N_{\rm cmb}}{61}\right)^{\!3} \Delta N,
\\
\label{zeta3a}
&& \tau_{\rm NL}^{\rm em} \simeq 0.4 \times 10^3 \, \vartheta^4 \left(\frac{N_{\rm cmb}}{61}\right)^{\!4} \Delta N,
\end{eqnarray}
respectively. Taking into account Eqs.~(\ref{PemPinf}), (\ref{zetao})-(\ref{zeta3o}),
and (\ref{zetaa})-(\ref{zeta3a}), we see that inflationary magnetogenesis is compatible with current
CMB data if
\begin{eqnarray}
\label{com1}
&& |\vartheta| \ll \frac{25}{(N_{\rm cmb}/61) (\Delta N)^{1/2}} \, , \\
\label{com2}
&& |\vartheta| \lesssim \frac{2}{(N_{\rm cmb}/61) (\Delta N)^{1/3}} \, , \\
\label{com3}
&& |\vartheta| \lesssim \frac{2}{(N_{\rm cmb}/61) (\Delta N)^{1/4}} \, ,
\end{eqnarray}
where the above three constraints come from the spectrum, bispectrum, and trispectrum
of the curvature perturbation, respectively. Since $\Delta N$
is a quantity greater than unity, we obtain that the stringent constraint on $|\vartheta|$
comes either form the bispectrum if $\Delta N \lesssim (25/2)^6 \simeq 4 \times 10^6$,
or from the spectrum otherwise.

Let us analyze the three cases in Eq.~(\ref{reality}), by assuming, for the sake of simplicity,
that the stringent constraint on $|\vartheta|$ comes form the bispectrum.

($i$) $\xi_1 = 0$. -- In this case (see section VI), $\langle (T_{\rm em})^\mu_\nu \rangle = 0$,
so that no curvature perturbations are generated by the inflation-produced electromagnetic field.

($ii$) $\xi_2 = 0$. -- In this case (see section VI),
$n = 1$, $\Upsilon_1 = 1$, and $\Upsilon_2 = 0$, which give $w_{\rm em} = 1/3$ and $\vartheta = 2$.
For $(N_{\rm cmb}/61) (\Delta N)^{1/3}$ of order unity,
the case $\xi_2 = 0$ is then compatible with current CMB data on curvature perturbations,
while for greater values it is excluded.

($iii$) $\xi_1/\xi_2 \in \mathbb{X}$. -- If $(N_{\rm cmb}/61) (\Delta N)^{1/3}$ is of order unity,
we have $n^5 |2 \Upsilon_1 - \Upsilon_2| = |\vartheta| \lesssim 2$. In this case,
looking at Fig.~1 and remembering the discussion at the end of section VI, we conclude that,
with the exclusion of those values very close to the boundary
$\partial \mathbb{X} = \{-1/3\} \cup \{1-2/\sqrt{3}\} \cup \{1\} \cup \{1+2/\sqrt{3}\}$,
the inflation-produced electromagnetic field (whose magnetic component directly accounts for cosmic magnetic fields)
generates curvature perturbations compatible with
CMB data. The regions in the parameter space $(\xi_1,\xi_2)$ with acceptable curvature perturbations,
instead, progressively shrink for increasing values of $(N_{\rm cmb}/61) (\Delta N)^{1/3}$,
as it appears from Fig.~3. Here, in the light gray regions the strength of the
actual, scale-invariant magnetic field is in the range $10^{-13} \G \leq B_0 \leq 5 \times 10^{-12} \G$
with no constraints on curvature perturbations imposed,
while in the shrunk dark gray regions the constraint~(\ref{com3}) has been imposed,
[the darkness increases as $(N_{\rm cmb}/61) (\Delta N)^{1/3}$ increases].


\begin{figure}
\begin{center}
\hspace{-0.5cm}
\includegraphics[scale=0.37,bb=0 0 600 600]{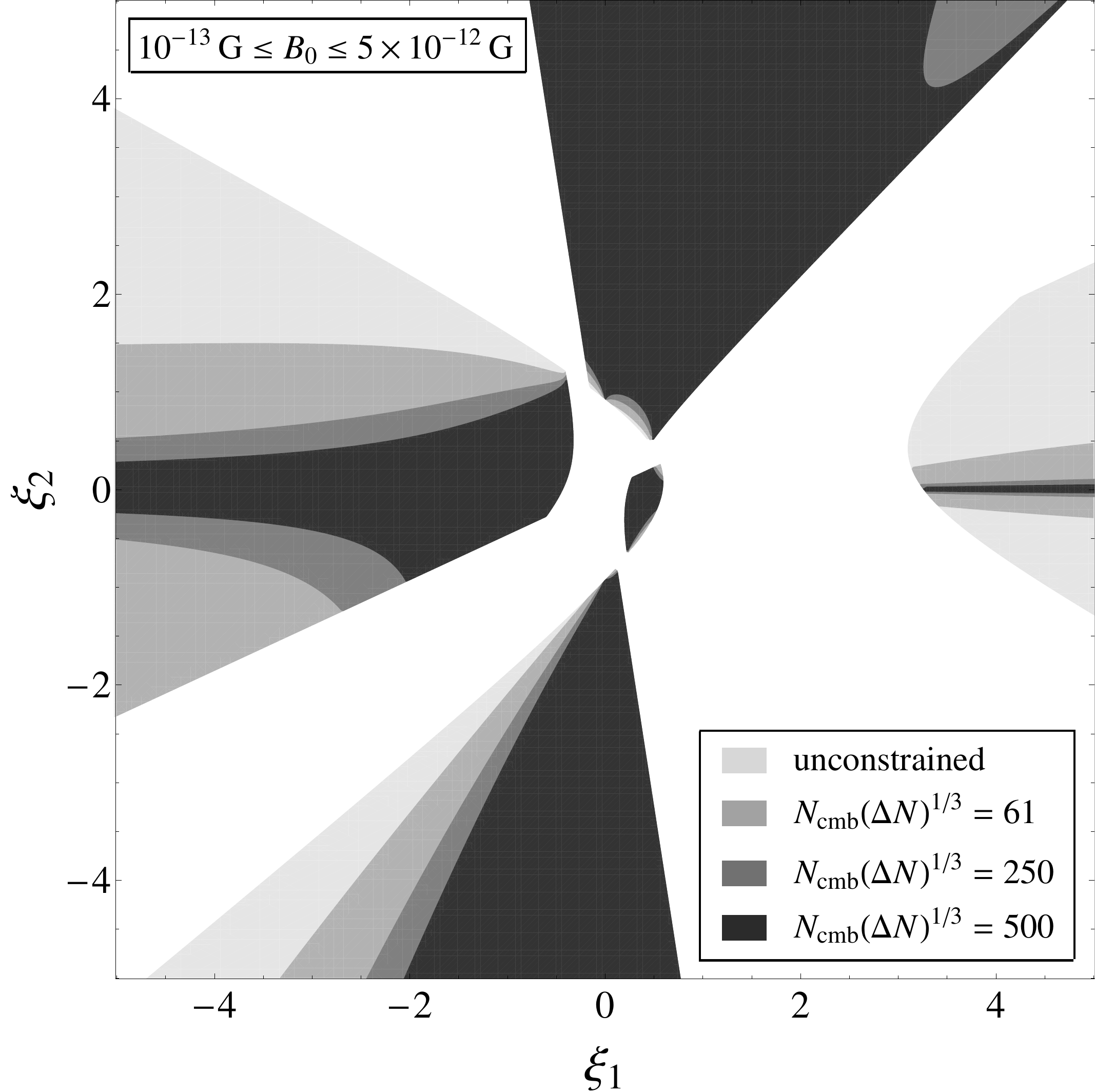}
\caption{Regions in the parameter space $(\xi_1,\xi_2)$, with $\xi_1/\xi_2 \in \mathbb{X}$ [see Eq.~(\ref{realityX})],
where $10^{-13} \G \leq B_0 \leq 5 \times 10^{-12} \G$ and the electromagnetic backreaction on
inflation is completely negligible, with no constraints on curvature perturbations imposed (light gray areas).
The dark gray regions represent the same regions after imposing the constraint~(\ref{com2})
which assures that scalar curvature perturbations generated by the inflation-produced electromagnetic field
are compatible with CMB observations [the darkness increases as $(N_{\rm cmb}/61) (\Delta N)^{1/3}$ increases].}
\end{center}
\end{figure}


\subsection{IXc. Tensor modes}

The inflation-produced electromagnetic
field affects, besides the scalar part of metric perturbation, also its tensor part, namely, it
produces gravitational waves. However, 
we expect that if $\mathcal{P}_\zeta^{\rm em} \ll \mathcal{P}_\zeta^{\rm inf}$,
then the same it is true for the gravitational wave spectra,
$\mathcal{P}_{\rm GW}^{\rm em} \ll \mathcal{P}_{\rm GW}^{\rm inf}$.
This is because, as discussed in~\cite{Barnaby}, the tensor modes are only produced gravitationally,
while the dominant source of the scalar modes is the direct coupling between the inflaton
and the electromagnetic field. In particular, the latter is enhanced with respect to the gravitational one by
a factor inversely proportional to the slow-roll parameter. Therefore, we expect that
\begin{equation}
\label{tensor} \frac{\mathcal{P}_{\rm GW}^{\rm em}}{\mathcal{P}_{\rm GW}^{\rm inf}} \sim
\epsilon \, \frac{\mathcal{P}_{\zeta}^{\rm em}}{\mathcal{P}_{\zeta}^{\rm inf}}
\end{equation}
on physical grounds. Accordingly, due to the assumption~(\ref{PemPinf}), we can safely
neglect the possible constraints on $\xi_1$ and $\xi_2$ coming from the production of
gravitational waves in our model of inflationary magnetogenesis.

\section{X. Discussion}

So far, we have not distinguished between explicit (not dynamical) and spontaneous
(dynamical) Lorentz violation. In the model at hand (described by a background rank-2 tensor),
the simplest realization of spontaneous
Lorentz violation is realized when a dynamical tensor operator $(\mathcal{K}_F)_{\mu\nu}$
is coupled to the electromagnetic field and it
acquires a vacuum expectation value different from zero,
$(k_F)_{\mu\nu} = \langle 0| (\mathcal{K}_F)_{\mu\nu} |0\rangle$.
In this case, the coupling between the rank-2 tensor field and
the electromagnetic field would be described by Lagrangian~(\ref{Lagrangian2})
with $(k_F)_{\mu\nu}$ replaced by $(\mathcal{K}_F)_{\mu\nu}$.

Decomposing $(\mathcal{K}_F)_{\mu\nu}$ in a classical part, $(k_F)_{\mu\nu}$,
and a quantum part, $\delta (k_F)_{\mu\nu}$, to wit, writing
\begin{equation}
\label{class+quant} (\mathcal{K}_F)_{\mu\nu} = (k_F)_{\mu\nu} + \delta (k_F)_{\mu\nu},
\end{equation}
all the above analysis remains valid only if the quantum fluctuations $\delta (k_F)_{\mu\nu}$
are dynamically negligible.
Assuming homogeneity and isotropy, we can write
$\delta (k_F)^\mu_\nu = \mbox{diag}(\delta \rho_K,-\delta p_K,-\delta p_K,-\delta p_K)$.
All the arguments and results in section~VI are then preserved if, roughly speaking,
\begin{equation}
\label{delta1} |\delta \rho_K| \, , |\delta p_K| \ll (-k\eta)^2  \left(\frac{\mPl}{M}\right)^{\!4} \! .
\end{equation}
In this case, in fact, the contribution of the quantum fluctuations of $(\mathcal{K}_F)_{\mu\nu}$
to the electromagnetic energy-momentum tensor are negligible in the scaling-invariant case
[see Eqs.~(\ref{T2new})-(\ref{T3new}) and Eqs.~(\ref{P3})-(\ref{P4})].
Although the condition~(\ref{delta1}) seems to be very restrictive, it could
be realized if, for example, $\delta (k_F)^\mu_\nu \sim (-k\eta)^\alpha$, with $\alpha \gtrsim 2$.

Finally, let us observe that the dynamics of the field $(\mathcal{K}_F)_{\mu\nu}$
is described by a Lagrangian of the type
$\mathcal{L}_{\mathcal{K}} = \mathcal{L}_{\mathcal{K}, \rm kin} + \mathcal{L}_{\mathcal{K}, \rm int}$,
where $\mathcal{L}_{\mathcal{K}, \rm kin}$ is the kinetic term
(whose involved expression, for the case of de Sitter spacetime, can be found in~\cite{Fronsdal}),
while the interaction term $\mathcal{L}_{\mathcal{K}, \rm int}$
contains, besides self-coupling terms,
all the couplings to other fields, included those with the photon and the inflaton.
In a complete and self-consistent model, which is beyond the aim of this paper,
one should also consider such a dynamics and consistently check that
the field $(\mathcal{K}_F)_{\mu\nu}$ does not appreciably back-react on the dynamics of inflation
and it generates curvature perturbations in agreement with CMB results.

\section{XI. Conclusions}

Astrophysical observations definitely indicate the
presence of microgauss magnetic fields in any type of galaxies,
and they give compelling indications of the existence of strong, large-scale
magnetic fields in galaxy clusters and cosmic voids.

A plausible hypothesis about their nature is that
they have a primordial origin. In particular, presently observed
magnetic fields could be nothing but primordial quantum
electromagnetic fluctuations excited during an inflationary epoch
of the universe which have survived until today.

However, since standard electrodynamics is conformally invariant,
large-scale magnetic fields cannot be generated in a conformally
invariant background spacetime, as a result of the Parker theorem.
Accordingly, whatever is the mechanism responsible
for generation of quantum electromagnetic fluctuations it must
break conformal invariance of Maxwell theory.

In this paper, we have analyzed the generation of cosmic magnetic
fields during (de-Sitter) inflation in a non-conformally invariant,
Lorentz-violating effective model of electrodynamics.
We have considered a Lorentz-violating extension of the kinetically
coupled scenario for magnetogenesis, where the latter is described by
a Lagrangian of the form $\mathcal{L}_{\rm em} = f(\phi) \mathcal{L}_M$. Here,
$\mathcal{L}_M$ is the Maxwell Lagrangian and $f(\phi)$ is a generic
coupling function between the photon and the inflaton $\phi$.
Lorentz violation is introduced in our model by considering the
Lagrangian $\mathcal{L}_{\rm em} = f(\phi) (\mathcal{L}_M + \mathcal{L}_{LV})$,
where $\mathcal{L}_{LV}$ incorporates
all possible Lorentz-violating renormalizable operators.

We have restricted our analysis to the case where Lorentz symmetry breaking
is implemented by the presence of a classical, homogeneous, and time-independent rank-2 symmetric
background tensor. Working in the weak-coupling regime, we have shown that the
creation of inflationary magnetic fields in this model proceeds similarly to the
case of magnetogenesis in the standard kinetically coupled scenario. The key difference
is that the new degrees of freedom represented by the components of the background tensor
can be tuned in such a way to suppress the electric part of the inflation-produced
electromagnetic energy-momentum tensor. This allows us to have, in de Sitter inflation with scale $\sim 10^{16} \GeV$,
a self-consistent model of magnetogenesis where the inflation-produced electromagnetic field
($i$) does not appreciably back-react on inflation,
($ii$) it does not significantly affect the primordial spectrum of both scalar (Gaussian and non-Gaussian)
and tensor curvature perturbations,
and ($iii$) it evolves after inflation to give a strong, scaling-invariant magnetic field
that directly accounts for galactic magnetic fields.



\begin{thebibliography}{99}

\bibitem{Review1}              P.~P.~Kronberg,
                               Rept.\ Prog.\ Phys.\ {\bf 57}, 325 (1994).

\bibitem{Review2}              L.~M.~Widrow,
                               Rev.\ Mod.\ Phys.\ {\bf 74}, 775 (2002).

\bibitem{Review3}              M.~Giovannini,
                               Int.\ J.\ Mod.\ Phys.\ D {\bf 13}, 391 (2004).

\bibitem{Review4}              J.~P.~Vall\'{e}e, New\ Astr.\ Rev.\ {\bf 48}, 763 (2004).

\bibitem{Review5}              R.~Beck,
                               AIP Conf.\ Proc.\ {\bf 1381}, 117 (2011).

\bibitem{Review6}              A.~Kandus, K.~E.~Kunze and C.~G.~Tsagas,
                               Phys.\ Rept.\ {\bf 505}, 1 (2011).

\bibitem{Review7}              D.~Ryu, D.~R.~G.~Schleicher, R.~A.~Treumann, C.~G.~Tsagas and L.~M.~Widrow,
                               Space Sci.\ Rev.\ {\bf 166}, 1 (2012)
                               [arXiv:1109.4055 [astro-ph.CO]];
                               L.~M.~Widrow, D.~Ryu, D.~R.~G.~Schleicher, K.~Subramanian, C.~G.~Tsagas and R.~A.~Treumann,
                               Space Sci.\ Rev.\ {\bf 166}, 37 (2012)
                               [arXiv:1109.4052 [astro-ph.CO]].

\bibitem{Review8}              R.~Durrer and A.~Neronov,
                               Astron.\ Astrophys.\ Rev.\ {\bf 21}, 62 (2013)
                               [arXiv:1303.7121 [astro-ph.CO]].

\bibitem{Review9}              T.~Prokopec,
                               astro-ph/0106247 (unpublished);

\bibitem{Pakmor}               R.~Pakmor, F.~Marinacci and V.~Springel,
                               Astrophys.\ J.\ {\bf 783}, L20 (2014).

\bibitem{Fermi}                E.~Fermi,
                               Phys.\ Rev.\ {\bf 75}, 1169 (1949).

\bibitem{Birrell-Davies}       N.~D.~Birrell and P.~C.~W.~Davies,
                               {\it Quantum Fields in Curved Space}
                               (Cambridge University Press, New York, 1982).

\bibitem{Parker-Toms}          L.~Parker and D.~J.~Toms,
                               {\it Quantum Field Theory in Curved Spacetime: Quantized Fields and Gravity}
                               (Cambridge University Press, Cambridge, England, 2009).

\bibitem{Turner-Widrow}        M.~S.~Turner and L.~M.~Widrow,
                               Phys.\ Rev.\ D {\bf 37}, 2743 (1988).

\bibitem{Ratra}                B.~Ratra,
                               Astrophys.\ J.\ {\bf 391}, L1 (1992).

\bibitem{G2}                   F.~D.~Mazzitelli and F.~M.~Spedalieri,
                               Phys.\ Rev.\ D {\bf 52}, 6694 (1995).

\bibitem{G3}                   D.~Lemoine and M.~Lemoine,
                               Phys.\ Rev.\ D {\bf 52}, 1955 (1995).

\bibitem{G4}                   M.~Gasperini, M.~Giovannini and G.~Veneziano,
                               Phys.\ Rev.\ Lett.\ {\bf 75}, 3796 (1995).

\bibitem{G5}                   A.~C.~Davis and K.~Dimopoulos,
                               Phys.\ Rev.\ D {\bf 55}, 7398 (1997).

\bibitem{G6}                   M.~Giovannini and M.~E.~Shaposhnikov,
                               Phys.\ Rev.\ D {\bf 57}, 2186 (1998).

\bibitem{G7}                   A.~Berera, T.~W.~Kephart and S.~D.~Wick,
                               Phys.\ Rev.\ D {\bf 59}, 043510 (1999).

\bibitem{G8}                   M.~Giovannini,
                               Phys.\ Rev.\ D {\bf 62}, 123505 (2000);
                               Phys.\ Rev.\ D {\bf 61}, 087306 (2000);
                               Phys.\ Rev.\ D {\bf 64}, 061301 (2001);
                               hep-ph/0104214;
                               Phys.\ Lett.\ B {\bf 659}, 661 (2008);
                               Phys.\ Rev.\ D {\bf 88}, 083533 (2013).

\bibitem{G9}                   A.~L.~Maroto,
                               Phys.\ Rev.\ D {\bf 64}, 083006 (2001).

\bibitem{G10}                  K.~Dimopoulos,
                               astro-ph/0105488 (unpublished).

\bibitem{G11}                  B.~A.~Bassett, G.~Pollifrone, S.~Tsujikawa, and F.~Viniegra,
                               Phys.\ Rev.\ D {\bf 63}, 103515 (2001).

\bibitem{G12}                  K.~Dimopoulos, T.~Prokopec, O.~Tornkvist, and A.~C.~Davis,
                               Phys.\ Rev.\ D {\bf 65}, 063505 (2002).

\bibitem{G13}                  M.~Marklund, P.~K.~S.~Dunsby, M.~Servin, G.~Betschart and C.~Tsagas,
                               Class.\ Quant.\ Grav.\ {\bf 20}, 1823 (2003).

\bibitem{G14}                  K.~Bamba and J.~Yokoyama,
                               Phys.\ Rev.\ D {\bf 69}, 043507 (2004).

\bibitem{G15}                  G.~Betschart, P.~K.~S.~Dunsby and M.~Marklund,
                               Class.\ Quant.\ Grav.\ {\bf 21}, 2115 (2004).

\bibitem{G16}                  T.~Prokopec and E.~Puchwein,
                               Phys.\ Rev.\ D {\bf 70}, 043004 (2004).

\bibitem{G17}                  A.~Ashoorioon and R.~B.~Mann,
                               Phys.\ Rev.\ D {\bf 71}, 103509 (2005).

\bibitem{G18}                  C.~G.~Tsagas,
                               Phys.\ Rev.\ D {\bf 72}, 123509 (2005);
                               Class.\ Quant.\ Grav.\ {\bf 22}, 393 (2005).

\bibitem{G19}                  M.~R.~Garousi, M.~Sami and S.~Tsujikawa,
                               Phys.\ Lett.\ B {\bf 606}, 1 (2005).

\bibitem{G21}                  C.~G.~Tsagas and A.~Kandus,
                               Phys.\ Rev.\ D {\bf 71}, 123506 (2005).

\bibitem{G22}                  M.~M.~Anber and L.~Sorbo,
                               JCAP {\bf 0610}, 018 (2006).

\bibitem{G23}                  C.~Zunckel, G.~Betschart, P.~K.~S.~Dunsby and M.~Marklund,
                               Phys.\ Rev.\ D {\bf 73}, 103509 (2006).

\bibitem{G24}                  K.~Bamba and M.~Sasaki,
                               JCAP {\bf 0702}, 030 (2007).

\bibitem{G25}                  A.~Akhtari-Zavareh, A.~Hojati and B.~Mirza,
                               Prog.\ Theor.\ Phys.\ {\bf 117}, 803 (2007).

\bibitem{G26}                  K.~E.~Kunze,
                               Phys.\ Lett.\ B {\bf 623}, 1 (2005);
                               Phys.\ Rev.\ D {\bf 77}, 023530 (2008).

\bibitem{G27}                  L.~Campanelli, P.~Cea, G.~L.~Fogli, and L.~Tedesco,
                               Phys.\ Rev.\ D {\bf 77}, 043001 (2008);
                               Phys.\ Rev.\ D {\bf 77}, 123002 (2008).

\bibitem{G29}                  K.~Bamba,
                               JCAP {\bf 0710}, 015 (2007).

\bibitem{G30}                  K.~Bamba and S.~D.~Odintsov,
                               JCAP {\bf 0804}, 024 (2008).

\bibitem{G31}                  K.~Bamba, N.~Ohta and S.~Tsujikawa,
                               Phys.\ Rev.\ D {\bf 78}, 043524 (2008).

\bibitem{G32}                  K.~Bamba, S.~Nojiri and S.~D.~Odintsov,
                               Phys.\ Rev.\ D {\bf 77}, 123532 (2008).

\bibitem{G34}                  K.~Bamba, C.~Q.~Geng and S.~H.~Ho,
                               JCAP {\bf 0811}, 013 (2008).

\bibitem{Campanelli2}          L.~Campanelli,
                               Int.\ J.\ Mod.\ Phys.\ D {\bf 18}, 1395 (2009).

\bibitem{G34bis}               R.~Durrer, L.~Hollenstein and R.~K.~Jain,
                               JCAP {\bf 1103}, 037 (2011).

\bibitem{G35}                  C.~T.~Byrnes, L.~Hollenstein, R.~K.~Jain and F.~R.~Urban,
                               JCAP {\bf 1203}, 009 (2012).

\bibitem{G35bis}               J.~Beltran Jimenez and A.~L.~Maroto,
                               Phys.\ Rev.\ D {\bf 83}, 023514 (2011); see also
                               D.~N.~Vollick,
                               Phys.\ Rev.\ D {\bf 86}, 084057 (2012).

\bibitem{G36}                  S.~L.~Cheng, W.~Lee and K.~W.~Ng,
                               arXiv:1409.2656 [astro-ph.CO].

\bibitem{G37}                  K.~Bamba,
                               Phys.\ Rev.\ D {\bf 91}, 043509 (2015).

\bibitem{Dolgov}               A.~Dolgov,
                               Phys.\ Rev.\ D {\bf 48}, 2499 (1993).

\bibitem{Barrow}               J.~D.~Barrow and C.~G.~Tsagas,
                               Phys.\ Rev.\ D {\bf 77}, 107302 (2008)
                               [Erratum-ibid.\ D {\bf 77}, 109904 (2008)].

\bibitem{Barrow1}              J.~D.~Barrow and C.~G.~Tsagas,
                               Mon.\ Not.\ Roy.\ Astron.\ Soc.\ {\bf 414}, 512 (2011);
                               J.~D.~Barrow, C.~G.~Tsagas and K.~Yamamoto,
                               Phys.\ Rev.\ D {\bf 86}, 023533 (2012).

\bibitem{Barrow2}              Y.~Shtanov and V.~Sahni,
                               JCAP {\bf 1201}, 088 (2013);
                               C.~G.~Tsagas,
                               arXiv:1412.4806 [astro-ph.CO].

\bibitem{Campanelli}           L.~Campanelli,
                               Phys.\ Rev.\ Lett.\ 111, {\bf 061301} (2013).

\bibitem{Campanelli5}          R.~Durrer, G.~Marozzi and M.~Rinaldi,
                               Phys.\ Rev.\ Lett.\ {\bf 111}, 229001 (2013);
                               L.~Campanelli,
                               Phys.\ Rev.\ Lett.\ {\bf 111}, 229002 (2013);
                               P.~G.~Tinyakov and F.~R.~Urban,
                               arXiv:1309.2270 [astro-ph.CO];
                               I.~Agullo and J.~Navarro-Salas,
                               arXiv:1309.3435 [gr-qc];
                               I.~Agullo, J.~Navarro-Salas and A.~Landete,
                               Phys.\ Rev.\ D {\bf 90}, 124067 (2014).

\bibitem{Demozzi}              V.~Demozzi, V.~Mukhanov and H.~Rubinstein,
                               JCAP {\bf 0908}, 025 (2009).

\bibitem{Membiela}             F.~A.~Membiela,
                               Nucl.\ Phys.\ B {\bf 885}, 196 (2014).

\bibitem{Ferreira}             R.~J.~Z.~Ferreira, R.~K.~Jain and M.~S.~Sloth,
                               JCAP {\bf 1310}, 004 (2013);
                               JCAP {\bf 1406}, 053 (2014).

\bibitem{Caprini}              C.~Caprini and L.~Sorbo,
                               JCAP {\bf 1410}, 056 (2014).

\bibitem{Tasinato}             G.~Tasinato,
                               JCAP {\bf 1503}, 040 (2015).

\bibitem{Ferreira2}            R.~Z.~Ferreira and J.~Ganc,
                               arXiv:1411.5362 [astro-ph.CO].

\bibitem{Barnaby}              N.~Barnaby, R.~Namba and M.~Peloso,
                               Phys.\ Rev.\ D {\bf 85}, 123523 (2012).

\bibitem{Gambini}              R.~Gambini and J.~Pullin,
                               Phys.\ Rev.\ D {\bf 59}, 124021 (1999).

\bibitem{Kostelecky3}          V.~A.~Kostelecky and S.~Samuel,
                               Phys.\ Rev.\ D {\bf 39}, 683 (1989).


\bibitem{Dolag et al}          K.~Dolag, M.~Bartelmann and H.~Lesch,
                               Astron.\ Astrophys.\ {\bf 387}, 383 (2002).

\bibitem{Dolag}                K.~Dolag (private communication).

\bibitem{Cheng et al}          B.~Cheng, A.~V.~Olinto, D.~N.~Schramm and J.~W.~Truran,
                               Phys.\ Rev.\ D {\bf 54}, 4714 (1996).

\bibitem{BBN}                  L.~Campanelli,
                               Phys.\ Rev.\ D {\bf 84}, 123521 (2011).

\bibitem{Kahniashvili2}        T.~Kahniashvili, A.~G.~Tevzadze, S.~K.~Sethi, K.~Pandey and B.~Ratra,
                               Phys.\ Rev.\ D {\bf 82}, 083005 (2010).

\bibitem{Kahniashvili}         T.~Kahniashvili, Y.~Maravin, A.~Natarajan, N.~Battaglia and A.~G.~Tevzadze,
                               Astrophys.\ J.\  {\bf 770}, 47 (2013).

\bibitem{Giovannini 2}         M.~Giovannini,
                               Class.\ Quant.\ Grav.\ {\bf 23}, R1 (2006).

\bibitem{Planck3}              P.~A.~R.~Ade {\it et al.}  [Planck Collaboration],
                               arXiv:1502.01594 [astro-ph.CO].

\bibitem{ellipsoidal}          L.~Campanelli, P.~Cea and L.~Tedesco,
                               Phys.\ Rev.\ Lett.\  {\bf 97}, 131302 (2006);
                               Phys.\ Rev.\ D {\bf 76}, 063007 (2007).

\bibitem{Kahniashvili3}        T.~Kahniashvili, G.~Lavrelashvili and B.~Ratra,
                               Phys.\ Rev.\ D {\bf 78}, 063012 (2008).

\bibitem{ionization}           J.~Chluba, D.~Paoletti, F.~Finelli and J.~A.~Rubino-Martin,
                               arXiv:1503.04827 [astro-ph.CO].

\bibitem{Kronberg et al}       P.~P.~Kronberg and P.~Simard-Normandin,
                               Nature {\bf 263}, 653 (1976).

\bibitem{Kronberg 2}           P.~P.~Kronberg,
                               AIP Conf.\ Proc.\ {\bf 558}, 451 (2001).

\bibitem{Neronov-Vovk}         A.~Neronov and I.~Vovk,
                               Science {\bf 328}, 73 (2010).

\bibitem{Tavecchio et al 1}    F.~Tavecchio, G.~Ghisellini, L.~Foschini, G.~Bonnoli, G.~Ghirlanda and P.~Coppi, 
                               Mon.\ Not.\ Roy.\ Astron.\ Soc.\ {\bf 406}, L70 (2010).

\bibitem{Tavecchio et al 2}    F.~Tavecchio, G.~Ghisellini, G.~Bonnoli and L.~Foschini,
                               Mon.\ Not.\ Roy.\ Astron.\ Soc.\ {\bf 414}, 3566 (2011).


\bibitem{Martin-Yokoyama}      J.~Martin and J.~Yokoyama,
                               JCAP {\bf 0801}, 025 (2008).

\bibitem{BICEP2}               P.~A.~R.~Ade, {\it et al.}  [BICEP2 Collaboration],
                               Phys.\ Rev.\ Lett.\ {\bf 112}, 241101 (2014).


\bibitem{Generation2a}         V.~A.~Kosteleck\'{y}, R.~Potting and S.~Samuel,
                               in {\it Proceedings of the 1991 Joint International Lepton-Photon
                               Symposium and Europhysics Conference on High Energy Physics},
                               edited by S.~Hegarty, K.~Potter, E.~Quercigh
                               (World Scientific, Singapore, 1992).

\bibitem{Generation2b}         O.~Bertolami and D.~F.~Mota,
                               Phys.\ Lett.\ B {\bf 455}, 96 (1999).

\bibitem{Generation2c}         A.~Mazumdar and M.~M.~Sheikh-Jabbari,
                               Phys.\ Rev.\ Lett.\ {\bf 87}, 011301 (2001).

\bibitem{Generation2d}         J.~Gamboa and J.~Lopez-Sarrion,
                               Phys.\ Rev.\ D {\bf 71}, 067702 (2005).

\bibitem{Generation2e}         K.~Bamba and J.~Yokoyama,
                               {\bf 70}, 083508 (2004).

\bibitem{Generation2f}         L.~Campanelli, P.~Cea and G.~L.~Fogli,
                               Phys.\ Lett.\ B {\bf 680}, 125 (2009).

\bibitem{Generation2g}         L.~Campanelli and P.~Cea,
                               Phys.\ Lett.\ B {\bf 675}, 155 (2009).

\bibitem{Generation2h}         L.~Campanelli,
                               Phys.\ Rev.\ D {\bf 80}, 063006 (2009);
                               Phys.\ Rev.\ D {\bf 90}, 105014 (2014).

\bibitem{Generation2i}         A.~P.~Kouretsis,
                               Eur.\ Phys.\ J.\ C {\bf 74}, 2879 (2014).

\bibitem{Kostelecky2}          V.~A.~Kostelecky,
                               Phys.\ Rev.\ D {\bf 69}, 105009 (2004).

\bibitem{Colladay}             D.~Colladay and V.~A.~Kostelecky,
                               Phys.\ Rev.\ D {\bf 58}, 116002 (1998).


\bibitem{Abel}                 W.~E.~Boyce and R.~C.~DiPrima,
                               {\it Elementary Differential Equations and Boundary Value Problems},
                               (Wiley, New York, 1986).

\bibitem{Data}                 V.~A.~Kostelecky and N.~Russell,
                               Rev.\ Mod.\ Phys.\ {\bf 83}, 11 (2011).




\bibitem{Campanelli3}          L.~Campanelli,
                               Eur.\ Phys.\ J.\ C {\bf 74}, 2690 (2014).

\bibitem{MHD1}                 A.~Brandenburg, K.~Enqvist and P.~Olesen,
                               Phys.\ Rev.\ D {\bf 54}, 1291 (1996).

\bibitem{MHD2}                 P.~Olesen,
                               Phys.\ Lett.\ B {\bf 398}, 321 (1997).

\bibitem{MHD3}                 D.~T.~Son,
                               Phys.\ Rev.\ D {\bf 59}, 063008 (1999).

\bibitem{MHD4}                 D.~Biskamp and W.~C.~M\"{u}ller,
                               Phys.\ Rev.\ Lett.\ {\bf 83}, 2195 (1999).

\bibitem{MHD5}                 M.~Christensson, M.~Hindmarsh and A.~Brandenburg,
                               Phys.\ Rev.\ E {\bf 64}, 056405 (2001);
                               Astron.\ Nachr.\ {\bf 326}, 393 (2005)
                               [astro-ph/0209119].

\bibitem{MHD6}                 A.~Brandenburg,
                               Science {\bf 292}, 2440, (2001).

\bibitem{MHD7}                 R.~Banerjee and K.~Jedamzik,
                               Phys.\ Rev.\ Lett.\ {\bf 91}, 251301 (2003)
                               [Erratum-ibid.\  {\bf 93}, 179901 (2004)];
                               Phys.\ Rev.\ D {\bf 70}, 123003 (2004).

\bibitem{MHD8}                 L.~Campanelli,
                               Phys.\ Rev.\ D {\bf 70}, 083009 (2004);
                               Phys.\ Rev.\ Lett.\ {\bf 98}, 251302 (2007).

\bibitem{MHD11}                T.~Kahniashvili, A.~Brandenburg, A.~G.~Tevzadze and B.~Ratra,
                               Phys.\ Rev.\ D {\bf 81}, 123002 (2010).

\bibitem{MHD12}                A.~G.~Tevzadze, L.~Kisslinger, A.~Brandenburg and T.~Kahniashvili,
                               Astrophys.\ J.\  {\bf 759}, 54 (2012).

\bibitem{MHD13}                T.~Kahniashvili, A.~G.~Tevzadze, A.~Brandenburg and A.~Neronov,
                               arXiv:1212.0596 [astro-ph.CO].

\bibitem{MHD14}                M.~Giovannini,
                               Phys.\ Rev.\ D {\bf 85}, 043006 (2012).

\bibitem{MHD15}                A.~Berera and M.~Linkmann,
                               Phys.\ Rev.\ E {\bf 90}, 041003 (2014).

\bibitem{MHD16}                A.~Brandenburg, T.~Kahniashvili and A.~G.~Tevzadze,
                               Phys.\ Rev.\ Lett.\  {\bf 114}, no. 7, 075001 (2015).

\bibitem{Campanelli4}          T.~Kahniashvili, A.~Brandenburg, L.~Campanelli, B.~Ratra and A.~G.~Tevzadze,
                               Phys.\ Rev.\ D {\bf 86}, 103005 (2012).

\bibitem{Dimopoulos-Davis}     K.~Dimopoulos and A.~C.~Davis,
                               Phys.\ Lett.\ B {\bf 390}, 87 (1997).

\bibitem{Kolb-Turner}          E.~W.~Kolb and M.~S.~Turner,
                               {\it The Early Universe}
                               (Addison-Wesley, Redwood City, California, 1990).

\bibitem{Olive}                K.~A.~Olive,
                               CERN Yellow Report CERN-2010-002, 149-196
                               [arXiv:1005.3955 [hep-ph]].


\bibitem{Hawking-Ellis}        S.~W.~Hawking and G.~F.~R.~Ellis,
                               {\it The large scale structure of space-time}
                               (Cambridge University Press, Cambridge, England, 1973).

\bibitem{Landau}               L.~D.~Landau and E.~M.~Lifshitz,
                               {\it The Classical Theory of Fields}
                               (Pergamon Press, Oxford, United Kingdom, 1971).


\bibitem{Suyama}               T.~Suyama and J.~Yokoyama,
                               Phys.\ Rev.\ D {\bf 86}, 023512 (2012).

\bibitem{Barnaby2}             M.~Shiraishi, E.~Komatsu, M.~Peloso and N.~Barnaby,
                               JCAP {\bf 1305}, 002 (2013).

\bibitem{Fujita2}              T.~Fujita and S.~Yokoyama,
                               JCAP {\bf 1309}, 009 (2013).

\bibitem{Lyth}                 D.~H.~Lyth, K.~A.~Malik and M.~Sasaki,
                               JCAP {\bf 0505}, 004 (2005).

\bibitem{Planck1}              P.~A.~R.~Ade {\it et al.}  [Planck Collaboration],
                               Astron.\ Astrophys.\ {\bf 571}, A24 (2014).

\bibitem{Planck0}              P.~A.~R.~Ade {\it et al.}  [Planck Collaboration],
                               Astron.\ Astrophys.\ {\bf 571}, A16 (2014).

\bibitem{Planck2}              P.~A.~R.~Ade {\it et al.}  [Planck Collaboration],
                               arXiv:1502.01592 [astro-ph.CO].

\bibitem{Feng}                 C.~Feng, A.~Cooray, J.~Smidt, J.~O'Bryan, B.~Keating and D.~Regan,
                               arXiv:1502.00585 [astro-ph.CO].


\bibitem{Fronsdal}             C.~Fronsdal,
                               Phys.\ Rev.\ D {\bf 20}, 848 (1979).


\end{thebibliography}
\end{document}